\documentclass[1p]{elsarticle}
\usepackage{mathtools}
\usepackage{amssymb}
\usepackage{lineno,hyperref}
\usepackage{commath}
\usepackage{bm}
\usepackage{siunitx}

\newcommand{\vect}[1]{\boldsymbol{\mathbf{#1}}}

\renewcommand{\Dif}[1]{\mathop{}\!\text{#1}}

\DeclareMathOperator*{\argmax}{arg\,max}

\journal{Journal of Sound and Vibration}

\biboptions{sort&compress}
\begin{document}
	
\begin{frontmatter}
		
\title{A General Theory for Bandgap Estimation in Locally Resonant Metastructures}
\author[gww]{C.\ Sugino}
\author[gww]{Y.\ Xia}
\author[gww]{S.\ Leadenham}
\author[gww,dg]{M.\ Ruzzene}
\author[gww]{A.\ Erturk\corref{cor1}}
\ead{alper.erturk@me.gatech.edu}

\address[gww]{G.W.\ Woodruff School of Mechanical Engineering, Georgia Institute of Technology, Atlanta, GA}
\address[dg]{D.\ Guggenheim School of Aerospace Engineering, Georgia Institute of Technology, Atlanta, GA}
\cortext[cor1]{Corresponding author}

\begin{abstract}
Locally resonant metamaterials are characterized by bandgaps at wavelengths that are much larger than the lattice size, enabling low-frequency vibration attenuation. Typically, bandgap analyses and predictions rely on the assumption of traveling waves in an infinite medium, and do not take advantage of modal representations typically used for the analysis of the dynamic behavior of finite structures. Recently, we developed a method for understanding the locally resonant bandgap in uniform finite metamaterial beams using modal analysis. Here we extend that framework to general (potentially non-uniform, 1D or 2D) locally resonant metastructures with specified boundary conditions using a general operator formulation. Using this approach, along with the assumption of an infinite number of resonators tuned to the same frequency, the frequency range of the locally resonant bandgap is easily derived in closed form. Furthermore, the bandgap expression is shown to be the same regardless of the type of vibration problem under consideration, depending only on the added mass ratio and target frequency. For practical designs with a finite number of resonators, it is shown that the number of resonators required for the bandgap to appear increases with the target frequency range, i.e. respective modal neighborhood. Furthermore, it is observed that there is an optimal, finite number of resonators which gives a bandgap that is wider than the infinite-absorber bandgap, and that the optimal number of resonators increases with target frequency and added mass ratio. As the number of resonators becomes sufficiently large, the bandgap converges to the derived infinite-absorber bandgap. Additionally, the derived bandgap edge frequencies are shown to agree with results from dispersion analysis using the plane wave expansion method. The model is validated experimentally for a locally resonant cantilever beam under base excitation. Numerical and experimental investigations are performed regarding the effects of mass ratio, non-uniform spacing of resonators, and parameter variations among the resonators.
\end{abstract}


\end{frontmatter}

\section{Introduction}

Inspired by photonic crystals in electromagnetism, researchers have long investigated phononic crystals for their potential to filter or redirect elastic waves \cite{laude}. Phononic crystals exhibit bandgaps (i.e.\ frequency ranges where elastic or acoustic waves cannot propagate) produced by Bragg scattering \cite{kushwaha, vasseur, hussein2006dispersive}, which occurs when the wavelength of the incident wave is on the order of the lattice constant of the crystal \cite{hussein,mace}. Therefore, a fundamental limitation of Bragg-based phononic crystals is that it is only possible to create low-frequency bandgaps using very large structures. In their seminal work, Liu et al.\ \cite{liu} showed the potential for locally resonant metamaterials to create bandgaps at wavelengths much larger than the lattice size, enabling the creation of low-frequency bandgaps in relatively small structures. Locally resonant metamaterials contain resonating elements, whether mechanical \cite{liu,  ho} or electromechanical \cite{casadei, jin, chen2013wave}, which are capable of storing and transferring energy. A significant body of research has examined locally resonant elastic/acoustic metamaterials of various types. Ho et al.\ \cite{ho} examined a similar system to Liu et al.\ \cite{liu} using a rigid frame with rubber-coated metal spheres as resonators. For that same type of system, Liu et al.\ \cite{liu2} found analytic expressions for the effective mass densities of 3D and 2D locally resonant metamaterials, showing that the effective mass becomes negative near the resonant frequency. Simplifying the analysis, others used a lumped-mass models to obtain the locally resonant bandgap \cite{wang, jensen}. Others have studied different implementations for resonators for different types of elastic waves \cite{yang,oudich,baravelli,zhu,matlack,nouh,wang2014}, and two-degree-of-freedom resonators \cite{yu}. Moving towards analytical predictions for the bandgap edge frequencies, Xiao et al.\ \cite{xiao2} used the plane wave expansion method to study flexural waves in a plate with periodically attached resonators, giving a method to predict the edges of the bandgap. Peng and Pai \cite{peng} also studied a locally resonant metamaterial plate, finding an explicit expression for the bandgap edge frequencies.

\begin{figure}
	\centering
	\includegraphics[width = 5in]{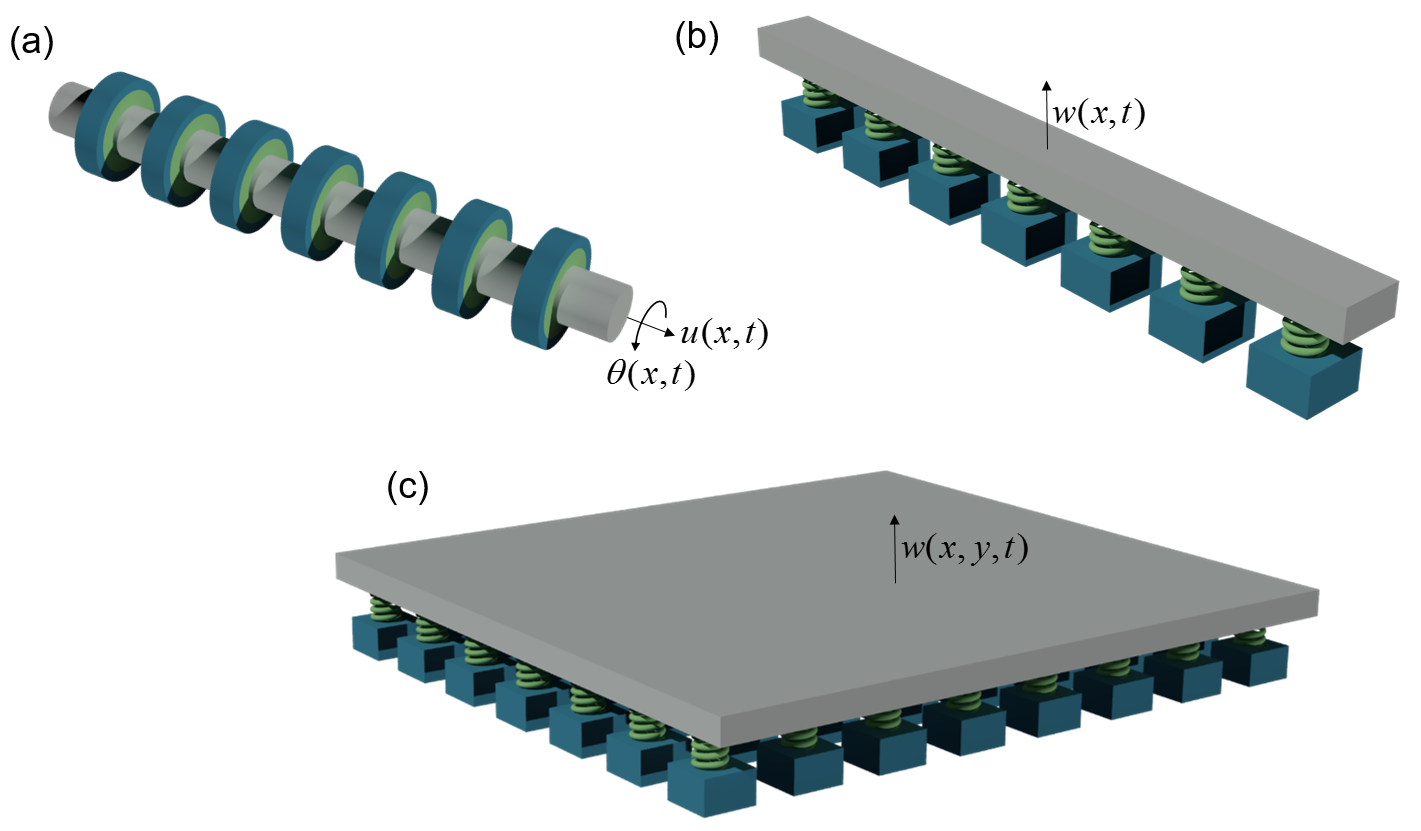}
	\caption{Examples of locally resonant metastructures: (a) Shaft/rod under torsional/longitudinal vibration, (b) beam under transverse vibration, and (c) plate under transverse vibration. Plain structure (primary structure) is shown in grey.\label{fig:sys_schem}}
\end{figure}

Much of the research on locally resonant metamaterials has relied on unit-cell based dispersion analysis, using techniques such as the plane wave expansion method to obtain the band structure of the metamaterial. This type of analysis lacks the information of modal behavior and cannot readily answer questions such as the dependence of the bandgap width on the number and spatial distribution of attachments in a finite structure. To this end we recently presented \cite{sugino2016mechanism} a modal analysis approach to bridge the gap between the lattice-based dispersion characteristics of locally resonant metamaterials and the modal interaction between a primary structure and its resonators. This paper extends the framework presented in \cite{sugino2016mechanism} for beams in bending to general (potentially non-uniform) 1D and 2D linear vibrating \textit{metastructures}, i.e. locally resonant metamaterial-based finite structures (Fig.\ \ref{fig:sys_schem}). A general form for the governing equations of the system is assumed using differential operator notation, and a modal expansion using the mode shapes of the structure without resonators provides significant simplification. Applying the assumption of an infinite number of resonators placed on the structure, the locally resonant bandgap edge frequencies are derived in closed form. This expression for the bandgap edge frequencies depends only on the resonant frequency of the resonators and the ratio of resonator mass to plain structure mass, and can be used for any typical vibrating structure (strings, rods, shafts, beams, membranes, or plates). To tie this work to other research in this field, the bandgap expression is validated numerically with the band structure obtained from the plane wave expansion method. To demonstrate that the derived bandgap expression is useful for design, we validate the bandgap expression for a finite number of uniformly distributed resonators. Additionally, we discuss the performance of metastructures with non-uniform distributions of resonators, as well as the effect of parameter variation in the resonant frequencies of the resonators. Finally, experimental validations are presented.

\section{Modal Analysis of Bandgap Formation}\label{sec:amm}

Consider a general partial differential equation governing the vibration of a forced, undamped, distributed parameter system of the form \cite{meirovitch}
\begin{equation}
\mathcal{L}\left[w(\vect{P},t)\right] + m(\vect{P})\ddot{w}(\vect{P},t) - \sum_{j=1}^Sk_ju_j(t)\delta(\vect{P} - \vect{P}_j) = f(\vect{P},t), \quad P \in D\label{eq:ge}
\end{equation}
with associated equations for the resonators
\begin{equation}
	m_j\ddot{u}_j(t) + k_ju_j(t) + m_j\ddot{w}(\vect{P}_j,t)  = 0, \quad j = 1,2,\ldots, S \label{ge2}
\end{equation}
where $w(\vect{P},t)$ is the displacement\footnote{The displacement $w(\vect{P},t)$ is assumed to be a scalar quantity for this derivation, but can be a vector. The units of the displacement field depend on the type of structure under consideration (Fig.\ \ref{fig:sys_schem}) and the respective operators and physical parameters are required to be consistent with that.} 
of a point with position vector $\vect{P}$ in the one or two-dimensional domain $D$, $\mathcal{L}$ is the linear homogeneous self-adjoint stiffness differential operator of order $2p$, where $p\ge1$ is an integer defining the order of the system, $m(\vect{P})$ is the mass density of the structure, an overdot indicates partial differentiation with respect to time, $k_j$ is the stiffness of the $j$th resonator, $m_j$ is the mass of the $j$th resonator, $u_j(t)$ is the displacement of the $j$th resonator, $\vect{P}_j$ is the attachment location of the $j$th resonator, $\delta(\vect{P})$ is the spatial Dirac delta function, $S$ is the total number of resonators, and $f(\vect{P},t)$ is the external force density. The choice of coordinate system depends on what is most convenient for the structure under consideration. Note that modal damping or resonator damping can easily be added at a later stage, but the bandgap phenomenon is most simply derived in the absence of any damping, without loss of generality. The boundary conditions can be written as
\begin{equation}
	\Dif{B}_iw(\vect{P},t) = 0, \quad i = 1, 2, \ldots, p, \quad \vect{P} \in \partial D \label{eq:bc}
\end{equation}
where $\Dif{B}_i$ are linear homogeneous boundary differential operators of order $[0, 2p-1]$ and $\partial D$ is the boundary of $D$. Equations\ \eqref{eq:ge} -\ \eqref{eq:bc} represent the general boundary value problem of a locally resonant metastructure.
Note that, because the force exerted by the resonators depends on the displacement of the structure, the mode shapes of the original structure without resonators (referred to as the ``plain structure" or the ``primary structure") are no longer the exact mode shapes of the full metastructure. Nevertheless, an expansion using the mode shapes of the plain structure provides significant simplification. It can be shown \cite{meirovitch} that the mode shapes of the plain structure satisfy the orthogonality conditions
\begin{gather}
	\int_Dm(\vect{P})\phi_r(\vect{P})\phi_s(\vect{P})\dif D = \delta_{rs}, \quad r,s = 1,2, \ldots\label{eq:orth1}\\
	\int_D\phi_r(\vect{P})\mathcal{L}\left[\phi_s(\vect{P})\right]\dif D = \omega_r^2\delta_{rs}, \quad r,s = 1,2,\label{eq:orth2}\ldots
\end{gather}
where $\phi_r(P)$ and $\phi_s(P)$ are the $r$th and $s$th normalized mode shapes of the plain structure, $\delta_{rs}$ is the Kronecker delta, and $\omega_r$ is the resonant frequency of the $r$th mode of the plain structure. Consider an approximate solution to the forced vibration problem of the form
\begin{equation}
	w(P,t) = \sum_{r=1}^N\eta_r(t)\phi_r(\vect{P}) \label{eq:amm}
\end{equation}
where $N$ is the number of modes to use in the expansion, and $\eta_r(t)$ is the modal weighting of the $r$th mode of the plain structure. When the domain $D$ is two-dimensional, it is often more natural to divide the modal summation into separate summations over the two modal indices. Nevertheless, it is always possible to rewrite the double summation as a single summation over all possible combinations of the two indices. Substituting Eq.\ \eqref{eq:amm} into Eq.\ \eqref{eq:ge} gives
\begin{equation}
	\sum_{r=1}^N\eta_r(t)\mathcal{L}\left[\phi_r(\vect{P})\right] + m(P)\sum_{r=1}^N\ddot{\eta}_r(t)\phi_r(\vect{P}) - \sum_{j=1}^Sk_ju_j(t)\delta(\vect{P} - \vect{P}_j)  = f(\vect{P},t), \quad \vect{P} \in D\label{eq:ge_exp}
\end{equation}
Multiplying Eq.\ \eqref{eq:ge_exp} by $\phi_s(P)$, integrating over the domain $D$, and applying the orthgonality conditions of the mode shapes gives
\begin{equation}
	\ddot{\eta}_r(t) + \omega_r^2\eta_r(t) - \sum_{j=1}^Sk_ju_j(t)\phi_r(\vect{P}_j) = q_r(t), \quad r = 1,2,\ldots, N\label{ge1}
\end{equation}
where
\begin{equation}
	q_r(t) = \int_Df(\vect{P},t)\phi_r\vect{P})\dif D
\end{equation}
Substituting Eq.\ \eqref{eq:amm} into Eq.\ \eqref{ge2} yields
\begin{equation}
	m_j\ddot{u}_j(t) + k_ju_j(t) + m_j\sum_{r=1}^N\ddot{\eta}_r(t)\phi_r(\vect{P}_j)  = 0, \quad j = 1,2,\ldots, S\label{ge4}
\end{equation}
To make the governing equations symmetric (i.e.\ coupled through inertial terms only), we can rewrite Eq.\ \eqref{ge1} by substituting Eq.\ \eqref{ge4} as
\begin{equation}
	\ddot{\eta}_r(t) + \omega_r^2\eta_r(t) + \sum_{j=1}^Sm_j\phi_r(\vect{P}_j)\sum_{k=1}^N\ddot{\eta}_k(t)\phi_k(\vect{P}_j) + \sum_{j=1}^Sm_j\ddot{u}_j(t)\phi_r(\vect{P}_j) = q_r(t), \quad r = 1,2,\ldots,N \label{ge3}
\end{equation}
Equations \eqref{ge4} and \eqref{ge3} form a system of $N + S$ coupled second order linear ordinary differential equations. Note that these equations must be altered slightly in the presence of base excitation, which is detailed in \ref{sec:base_motion}. It is possible to solve for the approximate mode shapes and resonant frequencies of the system, as well as the steady-state response to harmonic excitation. These methods provide little analytical insight, but they provide numerical solutions for a finite number of resonators. 

To gain more analytical insight, taking the Laplace transform of Eq.\ \eqref{ge2} gives
\begin{equation}
	U_j(s) = -\frac{s^2}{s^2 + \omega_{a,j}^2}\sum_{r=1}^NH_r(s)\phi_r(\vect{P}_j), \quad j = 1,2,\ldots, S \label{ujs}
\end{equation}
where $\omega_{a,j}^2 = k_j/m_j$ is the resonant frequency of the $j$th resonator. Taking the Laplace transform of Eq.\ \eqref{ge3} and substituting Eq.\ \eqref{ujs} yields
\begin{equation}
	(s^2 + \omega_r^2)H_r(s) + s^2\sum_{j=1}^S\frac{m_j\omega_{a,j}^2}{s^2 + \omega_{a,j}^2}\phi_r(\vect{P}_j)\sum_{k=1}^NH_k(s)\phi_k(\vect{P}_j) = Q_r(s), \quad r = 1,2,\ldots,N\label{scoupled}
\end{equation}
At this point, it is assumed that the resonators all have the same resonant frequency $\omega_{a,j} = \omega_t$. This is a strong assumption for a real physical implementation of a locally resonant metamaterial, since small parameter variations are always present. The effect of variations in $\omega_{a,j}$ will be discussed in Sec.\ \ref{subsec:param_unc}. To concretely define the mass of each attachment, divide the domain $D$ into $S$ distinct subdomains $D_j \subseteq D$ (analogous to unit cells, but not necessarily periodic), such that $\vect{P}_j \in D_j$, and define each mass $m_j$ as
\begin{equation}
m_j = \mu m(\vect{P}_j)\Delta D_j\label{massdistr}
\end{equation}
where 
\begin{equation}
	\Delta D_j = \int_{D_j}\dif D
\end{equation}
is the length or area of the subdomain $D_j$, and $\mu$ is the added mass ratio, defined as
\begin{equation}
	\mu = \frac{\displaystyle\sum_{j=1}^Sm_j}{\displaystyle\int_Dm(\vect{P})\dif D}
\end{equation}
which is the ratio of the total mass of the resonators to the total mass of the plain structure. The assumption in Eq.\ \eqref{massdistr} means that, as the number of resonators becomes large, the mass distribution of resonators will be proportional to the mass distribution of the plain structure by the mass ratio $\mu$. Substituting Eq.\ \eqref{massdistr} into Eq.\ \eqref{scoupled} gives
\begin{equation}
	(s^2 + \omega_r^2)H_r(s) + \mu\frac{s^2\omega_{t}^2}{s^2 + \omega_{t}^2}\sum_{j=1}^Sm(\vect{P}_j)\Delta D_j\phi_r(\vect{P}_j)\sum_{k=1}^NH_k(s)\phi_k(\vect{P}_j) = Q_r(s), \quad r = 1,2,\ldots,N \label{hsys}
\end{equation}
Equation \eqref{hsys} represents a system of $N$ coupled linear equations that cannot be solved for a simple analytical expression for $H_r(s)$ due to the coupling introduced by the presence of the absorbers. Again, it is possible at this stage to solve numerically for the response of a given system.

To simplify the system of equations given by Eq.\ \eqref{hsys}, consider what happens as $S\rightarrow \infty$ while keeping a fixed mass ratio $\mu$. The the regions $D_j$ become infinitesimal, the resonators become placed throughout the entire domain $D$, and we can say that
\begin{equation}
	\lim_{S\rightarrow\infty} \sum_{j=1}^S\Delta D_jm(\vect{P}_j)\phi_s(\vect{P}_j)\phi_r(\vect{P}_j) = \int_Dm(\vect{P})\phi_s(\vect{P})\phi_r(\vect{P})\dif D = \delta_{rs}, \quad r,s = 1, 2, \ldots
\end{equation}
This simplification is only exact in the limit as $S\rightarrow \infty$, but can be used as good approximations so long as 
\begin{equation}
	\sum_{j=1}^S\Delta D_jm(\vect{P}_j)\phi_s(\vect{P}_j)\phi_r(\vect{P}_j) \approx \int_Dm(\vect{P})\phi_s(\vect{P})\phi_r(\vect{P})\dif D\label{eq:infapprox}
\end{equation}
This approximation is key to the formation of the bandgap in structures with a finite number of resonators, as will be discussed in Sec.\ \ref{subsec:finite_S}. Under this assumption, Eq.\ \eqref{hsys} becomes
\begin{equation}
	(s^2 + \omega_r^2)H_r(s) + \mu\frac{s^2\omega_{t}^2}{s^2 + \omega_{t}^2}\sum_{k=1}^NH_k(s)\int_Dm(\vect{P})\phi_r(\vect{P})\phi_k(\vect{P})\dif D = Q_r(s), \quad r = 1,2,\ldots,N
\end{equation}
Then, applying the orthogonality of the mode shapes given in Eq.\ \eqref{eq:orth1}, the system of equations decouples, yielding the transfer function
\begin{equation}
\frac{H_r(s)}{Q_r(s)} = \frac{1}{s^2\left(1 + \dfrac{\mu\omega_t^2}{s^2 + \omega_t^2}\right) + \omega_r^2}, \quad r = 1,2,\ldots,N \label{hrs}
\end{equation}
It is clear from Eq.\ \eqref{hrs} that the presence of the resonators adds a frequency-dependent mass term. The boundary conditions of the structure do not affect the form of Eq.\ \eqref{hrs}. Additionally, the exact form of the original vibration problem is irrelevant, suggesting that this analysis can be used for a wide variety of vibrating systems.

Similar simplifications are possible for the resonator displacements, which become continuous in space with the assumption of an infinite number of resonators. Substituting Eq.\ \eqref{hrs} into Eq.\ \eqref{ujs}, and replacing the discrete locations $\vect{P}_j$ with a continuous variable $\vect{P}$,
\begin{equation}
	U(\vect{P}, s) = -\frac{s^2}{s^2 + \omega_t^2}\sum_{r=1}^N\frac{Q_r(s)\phi_r(\vect{P})}{s^2\left(1 + \dfrac{\mu\omega_t^2}{s^2 + \omega_t^2}\right) + \omega_r^2}\label{eq:u_cont}
\end{equation}
The displacement of the resonators can now be written as a linear combination of the mode shapes of the plain structure. Thus, in analogy with Eq.\ \eqref{eq:amm}, we can define the displacement of the resonators as
\begin{equation}
	u(\vect{P}, t) = \sum_{r=1}^N\psi_r(t)\phi_r(\vect{P})
\end{equation}
or, in the Laplace domain,
\begin{equation}
	U(\vect{P}, s) = \sum_{r=1}^N\Psi_r(s)\phi_r(\vect{P})\label{eq:amm_u}
\end{equation}
where, matching Eq. \eqref{eq:amm_u} to Eq.\ \eqref{eq:u_cont},
\begin{equation}
	\frac{\Psi_r(s)}{Q_r(s)} = \frac{-s^2}{s^2\left(s^2 + (1+\mu)\omega_t^2\right) + \omega_r^2(s^2+\omega_t^2)}, \quad r = 1,2,\ldots, N
\end{equation}

\begin{figure}
	\centering
	\includegraphics[width = 4in]{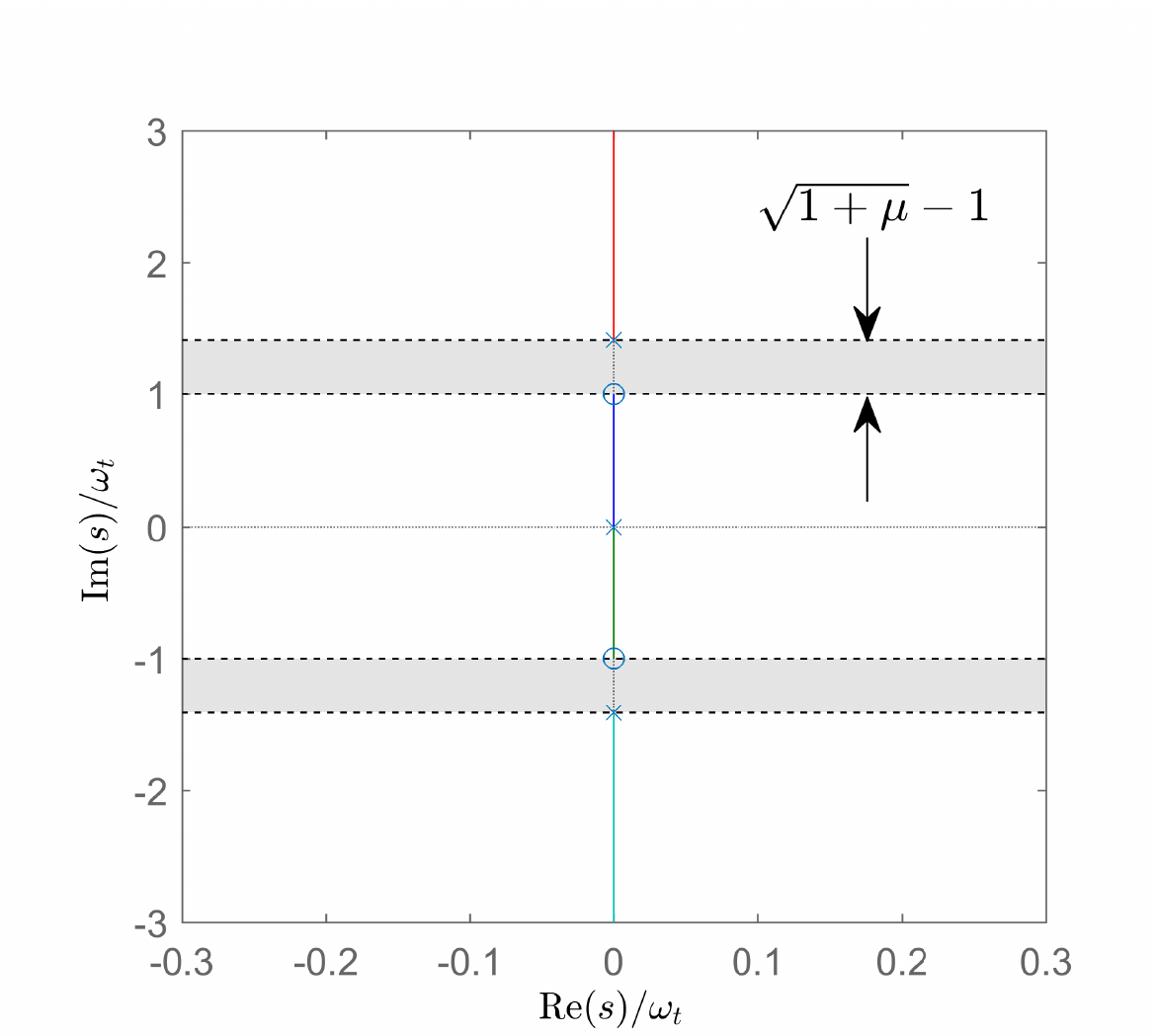}
	\caption{Root locus plot for the transfer function in Eq.\ \eqref{eq:plant} with gain $K = \omega_r^2$ showing the bandgap for $\mu = 1$. The markers correspond to poles (``x") and zeros (``o") at $\omega_r^2 = 0$. Solid lines show the root locus plot for $\omega_r^2 > 0$. The grey region bounded by the dashed lines shows the frequency range of the bandgap.\label{fig:rlocus}}
\end{figure}

The resonances of the primary structure associated with each mode shape are given by the imaginary part of the poles of the transfer function in Eq.\ \eqref{hrs}. To understand the behavior of these poles for various modes of the plain beam,  we can consider the modal response as the closed-loop transfer function of a feedback system with proportional gain $\omega_r^2$, with plant transfer function
\begin{equation}
	G(s) = \frac{1}{s^2\left(1 + \dfrac{\mu\omega_t^2}{s^2 + \omega_t^2}\right)} = \frac{s^2 + \omega_t^2}{s^2\left(s^2 + (1 + \mu)\omega_t^2\right)}\label{eq:plant}
\end{equation}
Under this interpretation, the bandgap edge frequencies can be obtained geometrically using well-known root locus analysis \cite{ogata}. Specifically, there are zeros at $s = \pm j\omega_t$, poles at $s = \pm j\omega_t\sqrt{1 + \mu}$, and a second order pole at $s = 0$. Since poles must go to infinity or to zeros as $\omega_r^2\rightarrow\infty$, and all of the poles and zeros are on the imaginary axis, it is clear that there can be no poles within the range $\omega_t < \text{Im}(s) < \omega_t\sqrt{1 + \mu}$, as shown in Fig.\ \ref{fig:rlocus}. Thus, the frequency range
\begin{equation}
	\boxed{\omega_t < \omega < \omega_t\sqrt{1 + \mu}}\label{eq:bg}
\end{equation} 
defines the infinite-absorber locally resonant bandgap. This is more intuitively understood as the frequency range in which the effective dynamic mass of the system becomes negative. This result agrees with the bandgap edge frequencies found in \cite{peng}, which were obtained for flexural waves in a plate using unit-cell dispersion analysis. The zero at $s = j\omega_t$  is an antiresonance that is present for every mode shape of the structure, resulting in zero displacement everywhere on the structure. The \textit{bandwidth} of the bandgap is then
\begin{equation}
	\Delta \omega_\infty = \omega_t\left(\sqrt{1 + \mu} - 1\right)\label{eq:bg_width}
\end{equation}

The root locus interpretation of the bandgap is well-suited for the analysis of general linear attachments. Furthermore, this method can be used to understand how to generate multiple bandgaps, simply by correctly placing poles and zeros in the plant transfer function, and extracting the required response/force relationship for the resonators. 

From this analysis, it is clear that each mode shape has two associated resonant frequencies, one below the bandgap frequency range and one above, corresponding to the two branches of the root locus with $\text{Im}(s) > 0$. These resonances correspond to different This can be seen by relating the modal weightings of the resonators and structure:
\begin{equation}
\frac{\Psi_r(j\omega)}{H_r(j\omega)} = \frac{\omega^2}{\omega_t^2 - \omega^2}, \quad r = 1,2,\ldots,N
\end{equation}
Thus, $\Psi_r$ and $H_r$ have the same signs for $\omega < \omega_t$, and opposite signs for $\omega > \omega_t$. Since this is true for all $r$, the structure and resonators are always in phase for $\omega < \omega_t$, and always out of phase for $\omega > \omega_t$.

To obtain an exact expression for the new resonant frequencies of each mode, assume an excitation frequency $\omega$ and substitute $s = j\omega$ into Eq.\ \eqref{hrs}, giving
\begin{equation}
	\frac{H_r(j\omega)}{Q_r(j\omega)} = \frac{1}{-\omega^2\left(1 + \dfrac{\mu\omega_t^2}{\omega_t^2 - \omega^2}\right) + \omega_r^2}, \quad r = 1,2,\ldots,N \label{hjw}
\end{equation} 
The roots of the denominator give the new resonant frequencies associated with each mode. Solving yields the two positive real roots as
\begin{align}
	\hat{\omega}^+ &= \sqrt{\dfrac{1 + \mu + \Omega_r^2}{2}\left(1 + \sqrt{1 - \dfrac{4\Omega_r^2}{(1 + \mu + \Omega_k^2)^2}}\right)}\\
	\hat{\omega}^- &= \sqrt{\dfrac{1 + \mu + \Omega_r^2}{2}\left(1 - \sqrt{1 - \dfrac{4\Omega_r^2}{(1 + \mu + \Omega_k^2)^2}}\right)}.
\end{align}
where $\hat{\omega} = \omega/\omega_t$ and $\Omega_r = \omega_r/\omega_t$ are dimensionless frequencies.  From these expressions, the bandgap edge frequencies can be obtained by considering the values of $\hat{\omega}$ at $\Omega_r = 0$ and $\Omega_r \rightarrow\infty$, giving the same result as Eq.\ \eqref{eq:bg}.

The boundary or stiffness operators from the original boundary value problem did not play a role in this derivation, suggesting that the infinite-absorber bandgap can be placed at any frequency range, without regard for the natural dynamics of the structure. It is not necessary to know the resonances or mode shapes of the original structure; the only important property of the plain structure is its mass distribution, which must be taken into account when selecting the mass distribution for the resonators. This is extremely important for practical designs attempting to create locally resonant bandgaps, as many structures have non-uniform mass distributions. Additionally, the mass ratio $\mu$ determines the width of the bandgap, and must be increased to increase the bandgap width at a fixed target frequency $\omega_t$.

\section{Bandgap Comparison and Model Validation with Dispersion Analysis by Plane Wave Expansion}\label{sec:pwem}
\begin{figure}
	\centering
	\includegraphics[]{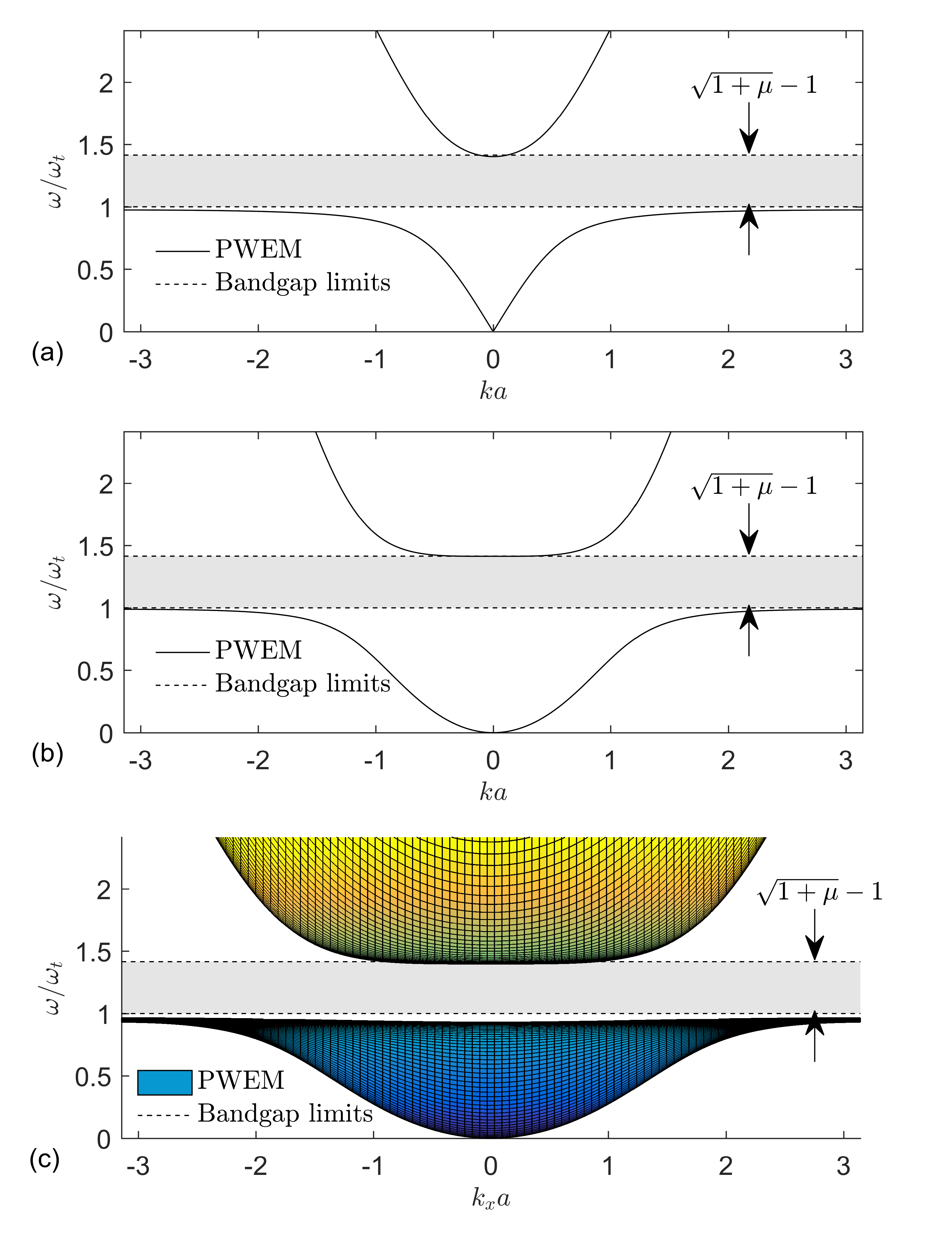}
	\caption{Dispersion curves computed using the PWEM for (a) longitudinal waves in a rod, (b) flexural waves in a beam, (c) flexural waves in a plate, with the expected bandgap shown in grey bounded by the dashed lines. Parameters $\mu = 1$, $f_t = \omega_t/(2\pi) = \SI{500}{Hz}$, $E = \SI{71}{GPa}$, thickness $h = \SI{2}{mm}$, $\rho = \SI{2700}{kg/m^3}$, $\nu = 0.3$, lattice constant for (a) $a = \SI{0.75}{m}$, lattice constant for (b) and (c) $a = \SI{0.05}{m}$.\label{pwem_comp}}
\end{figure}

The plane wave expansion method (PWEM) is commonly used to find the band structure of phononic crystals (for additional details see \ref{sec:pwem_app}, refer to \cite{laude, xiao2}). The key assumption is that the resonators are placed periodically on an infinite structure, and a Fourier series type expansion of plane waves is assumed for the amplitude of the response, i.e.
\begin{equation}
w(\vect{P}) = \sum_{m}W_1(\vect{G}_m)e^{-i(\vect{k} + \vect{G}_m)\cdot \vect{P}}
\end{equation}
where $\vect{G}_m$ are the reciprocal lattice vectors, $W_1$ are the plane wave amplitudes, and $\vect{k}$ is the Bloch wavevector. A similar expansion can be used for the material properties of the structure. By substituting this expansion into the governing equation for the structure, applying the periodicity of the resonator locations, multiplying by an exponential, and applying orthogonality, an eigenvalue problem is obtained. This eigenvalue problem must be solved for the eigenfrequencies $\omega$ at every Bloch wavevector in the irreducible region of the first Brillouin zone. It is possible to apply this technique to a general vibrating system whose governing equation is given in operator notation, which is described in more detail in \ref{sec:pwem_app}. For sufficiently small lattice constants, the bandgap from the PWEM dispersion analysis agree well with the edge frequencies in Eq.\ \eqref{eq:bg}, as shown in Fig.\ \ref{pwem_comp}  for various types of structures (Fig.\ \ref{fig:sys_schem}) in which the plain structure (i.e. structure in the absence of the resonators) exhibits 1D non-dispersive, 1D dispersive, and 2D dispersive behaviors.

\section{Numerical Studies}\label{sec:numstudies}
In this section, first we show that the infinite-absorber approximation Eq.\ \eqref{eq:infapprox} and the resulting bandgap expression Eq.\ \eqref{eq:bg} are useful for systems bearing a sufficient number of uniformly distributed resonators. From these numerical studies, a simple method is proposed to determine the optimal number of resonators for a given system. Next, since the Riemann sum approximation in Eq.\ \eqref{eq:infapprox} places no restrictions on the placement of resonators, non-uniform distributions of resonators are investigated. Finally, for practical implementations of locally resonant metastructures, the effect of small variations in the resonator natural frequencies is discussed.

For simplicity, the numerical studies presented here will focus on a cantilever beam excited by base motion. The associated mode shapes, resonant frequencies, and appropriate modal excitation expression for cantilever beams are given in \ref{subsec:beams}. Note that the analysis presented thus far is independent of the type of structure under consideration, and so these results can be extended easily to other vibrating systems. As an example, plate vibration will be discussed briefly, with the corresponding modal information in \ref{subsec:plates}. The stiffness operators and appropriate mass densities are given for the typical vibrating systems in \ref{sec:lmdefs}. Since the procedure for finding mode shapes given the boundary conditions of the system is well known, we refer to Meirovitch \cite{meirovitch} for any additional details.

\subsection{Finite Number of Uniformly Distributed Resonators}\label{subsec:finite_S}
The infinite-absorbers approximation in Eq.\ \eqref{eq:infapprox} that yields the simple bandgap expression in Eq.\ \eqref{eq:bg} is only exact as $S \rightarrow \infty$. For practical design purposes, it is important to understand whether Eq.\ \eqref{eq:bg} can be used in the case of finite $S$ as well as how the bandgap changes with a reduced number of attachments. It is also important to understand how the required number of attachments changes with changing target frequency (i.e.\ mode neighborhood). To limit the number of independent variables, consider a uniform cantilever beam of length $L$ with $S$ evenly spaced resonators, such that there is always a resonator at the tip of the cantilever, or $x_j = jL/S$. Since the cantilever is uniform, the mass of each resonator will be the same, given by $m_j = \mu m L / S$. Each resonator has the same natural frequency $\omega_t$. 

\begin{figure}
	\centering
	\includegraphics[]{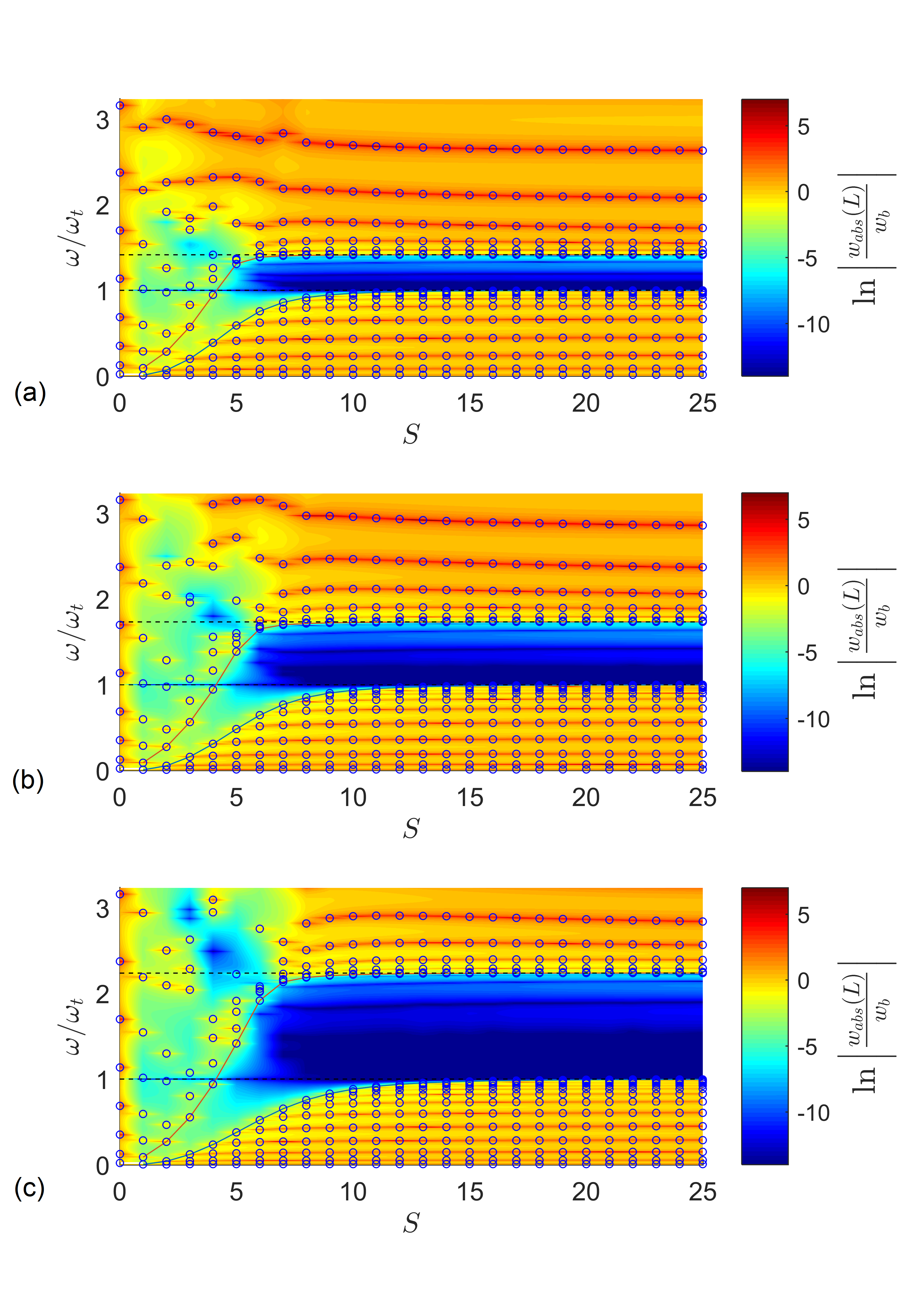}
	\caption{Resonances and transmissibility vs. number of resonators for $\omega_t = 50\omega_1$ with (a) $\mu = 1$, (b) $\mu = 2$, and (c) $\mu = 4$. Small circles indicate resonant frequencies, and the heatmap shows transmissibility on a log-scale. Dashed lines show the expected bandgap edge frequencies for sufficiently large numbers of attachments. Solid lines show $\omega_{S+1}$ and $\omega_S$. \label{fig:SvsOmega_mu}}
\end{figure}

\begin{figure}
	\centering
	\includegraphics[]{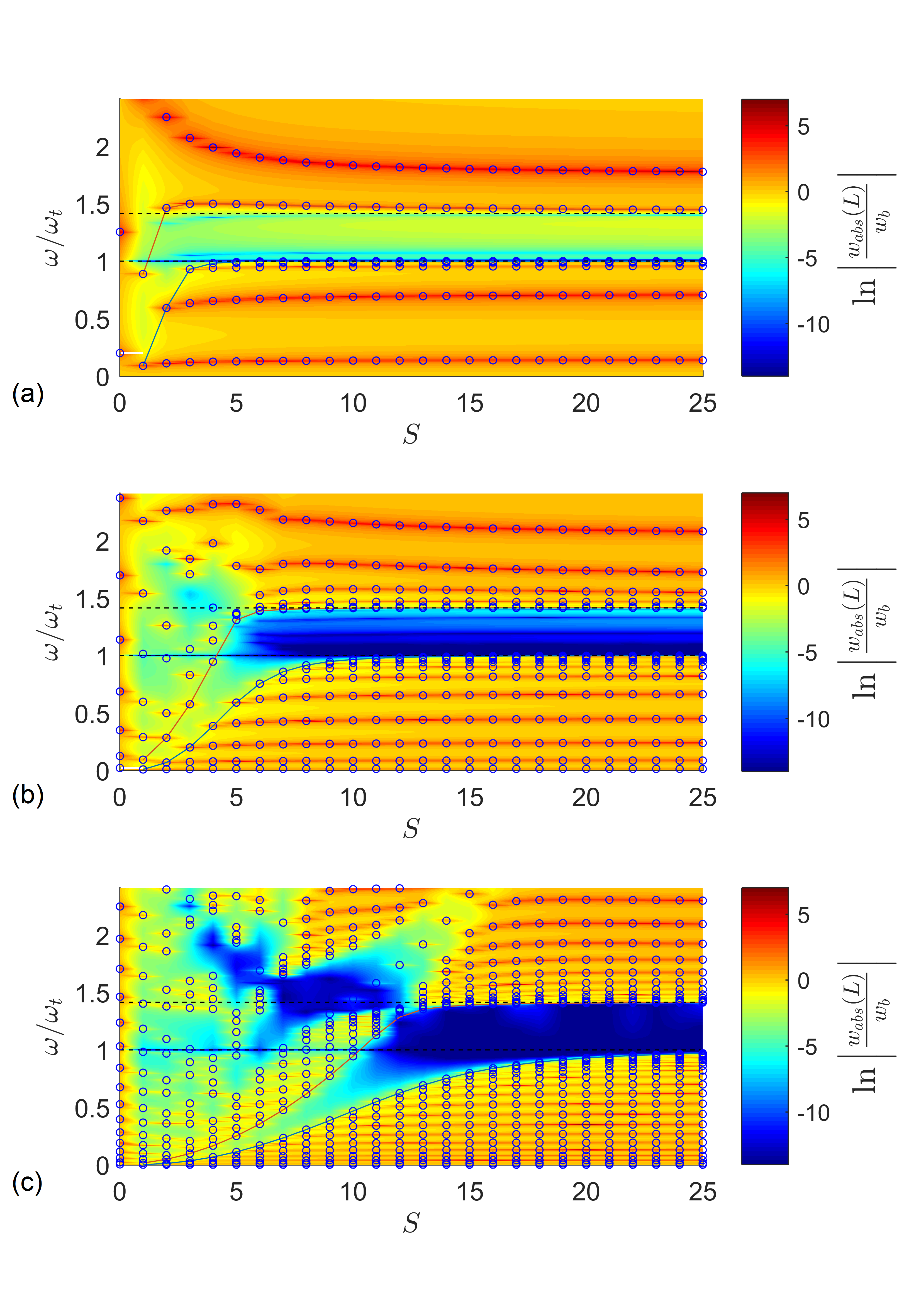}
	\caption{Resonances and transmissibility vs. number of resonators for $\mu = 1$, with (a) $\omega_t = 5\omega_1$, (b) $\omega_t = 50\omega_1$, and (c) $\omega_t = 300\omega_1$. Small circles indicate resonant frequencies, and the heatmap shows transmissibility on a log-scale. Dashed lines show the expected bandgap edge frequencies for sufficiently large numbers of attachments. Solid lines show $\omega_{S+1}$ and $\omega_S$.\label{fig:SvsOmega_wt}}
\end{figure}

Under these conditions, it is possible to find the resonant frequencies of the full system as a function of $S$ only using Eqs.\ \eqref{ge4} and \eqref{ge3}, yielding the resonant frequencies $\omega_i(S)$ for $i = 1,\ldots,N+S$. Similarly, the deformed shape of the beam can be found at each value of $S$ as a function of excitation frequency $\omega$. Sets of plots showing the resonant frequencies and transmissibility for the cantilever beam excited by base motion are shown in Fig.\ \ref{fig:SvsOmega_mu} (constant $\omega_t$, three values of $\mu$) and Fig.\ \ref{fig:SvsOmega_wt} (constant $\mu$, three values of $\omega_t$). 

\begin{figure}
	\centering
	\includegraphics[]{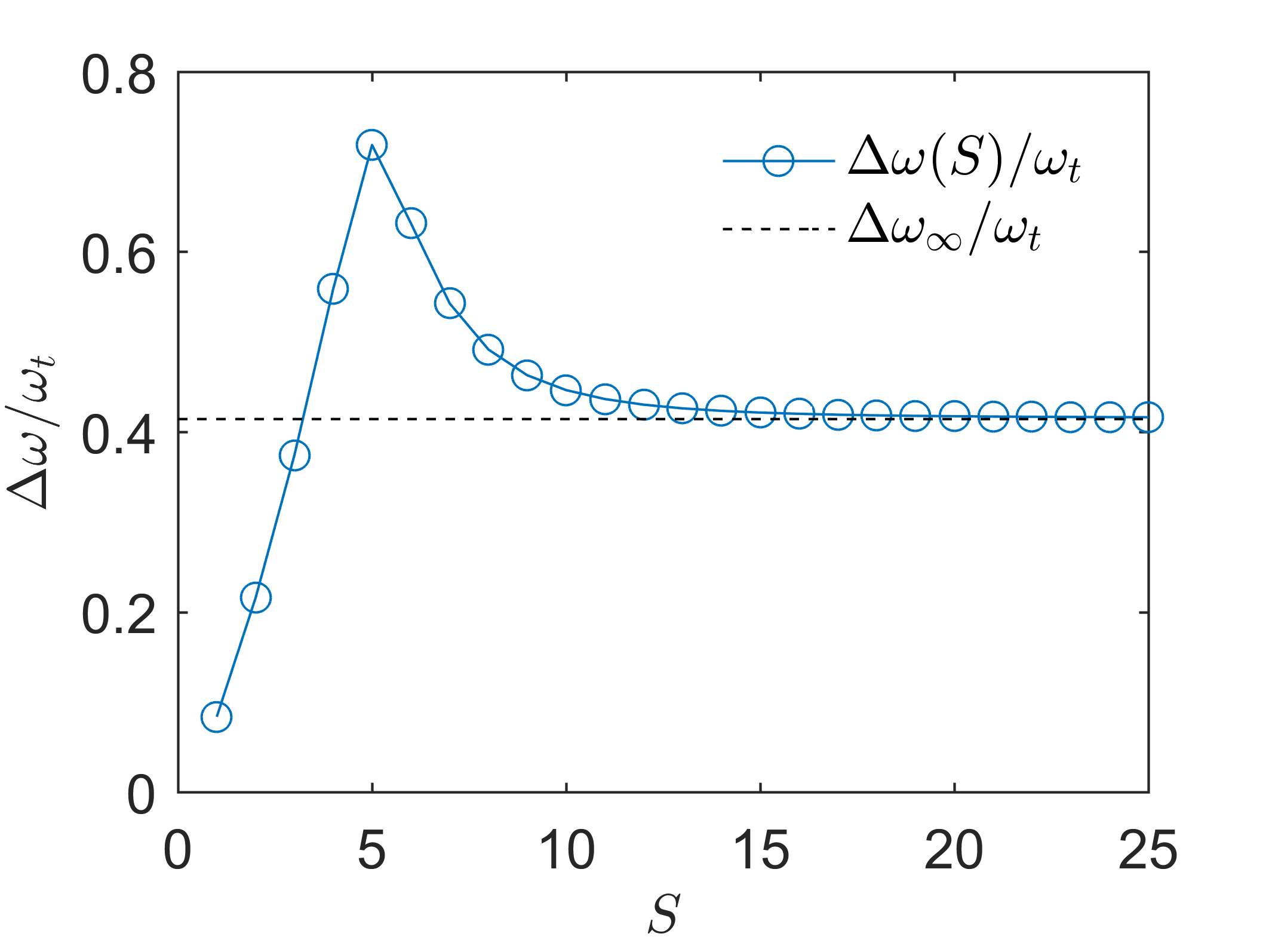}
	\caption{Effective bandgap width vs. number of resonators for a cantilever beam, $\omega_t = 50\omega_1$, $\mu = 1$. The maximum bandgap width is clearly given at $S_{opt} = 5$.\label{fig:bw_vs_S}}
\end{figure}

Both Figs.\ \ref{fig:SvsOmega_mu} and \ref{fig:SvsOmega_wt} contain critical information to understand the nature of bandgap formation in a finite structure. In all cases, the bandgap is clearly indicated by a sharp reduction in transmissibility. From Figs.\ \ref{fig:SvsOmega_mu} and \ref{fig:SvsOmega_wt}, it is clear that a certain minimum number of attachments is required for the formation of a bandgap. Remarkably, it is noted that the widest bandgap is not for the largest number of attachments (further discussed with Fig.\ \ref{fig:bw_vs_S}). However, there is a clear convergence to a fixed bandgap width that is accurately estimated by Eq.\ \eqref{eq:bg_width} (limits of which are given by the dashed lines in Figs.\ \ref{fig:SvsOmega_mu} and \ref{fig:SvsOmega_wt}). Figure \ref{fig:SvsOmega_wt} shows that more attachments (larger S value) are required for bandgap formation if one targets a relatively high frequency neighborhood (i.e. higher mode) of the main structure. It can be expected that these plots would shift for different combinations of boundary conditions and types of vibrating structures, especially for low $S$ values, because the mode shapes will be different.

As shown by the solid lines in Fig.\ \ref{fig:SvsOmega_mu} and \ref{fig:SvsOmega_wt}, it is clear that the bandgap forms between the resonant frequencies $\omega_S$ and $\omega_{S+1}$ as $S$ becomes large. Thus, for a given value of $S$, we can define the effective bandgap width as
\begin{equation}
	\Delta \omega (S) = \omega_{S+1}(S) - \omega_S(S)
\end{equation}
where it is clear that
\begin{equation}
	\lim_{S\rightarrow\infty}\Delta\omega(S) = \Delta \omega_\infty = \omega_t\left(\sqrt{1 + \mu} - 1\right)
\end{equation}
Note that for $\Delta\omega$ to be a valid bandgap width, we must have $\omega_{S+1} > \omega_t$, since the antiresonance at $\omega_t$ is an integral part of the bandgap. Interestingly, the bandgap width predicted by Eq.\ \eqref{eq:bg_width} is generally \emph{not} the maximum value of $\Delta\omega(S)$. Consequently, we can find a finite value $S_{opt}$ which maximizes $\Delta\omega(S)$ for a particular structure, or
\begin{align}
	S_{opt} = \argmax_{\substack{S \in \mathbb{Z}_{\ge 1}}}{} & \Delta \omega(S)\\
	\text{s.t. } & \omega_{S+1} > \omega_t\nonumber
\end{align}
A plot showing $\Delta\omega(S)$ vs.\ $S$ is shown in Fig.\ \ref{fig:bw_vs_S}, revealing the clearly defined maximum bandgap width for a particular set of parameters.

\begin{figure}
	\centering
	\includegraphics[width = 5in]{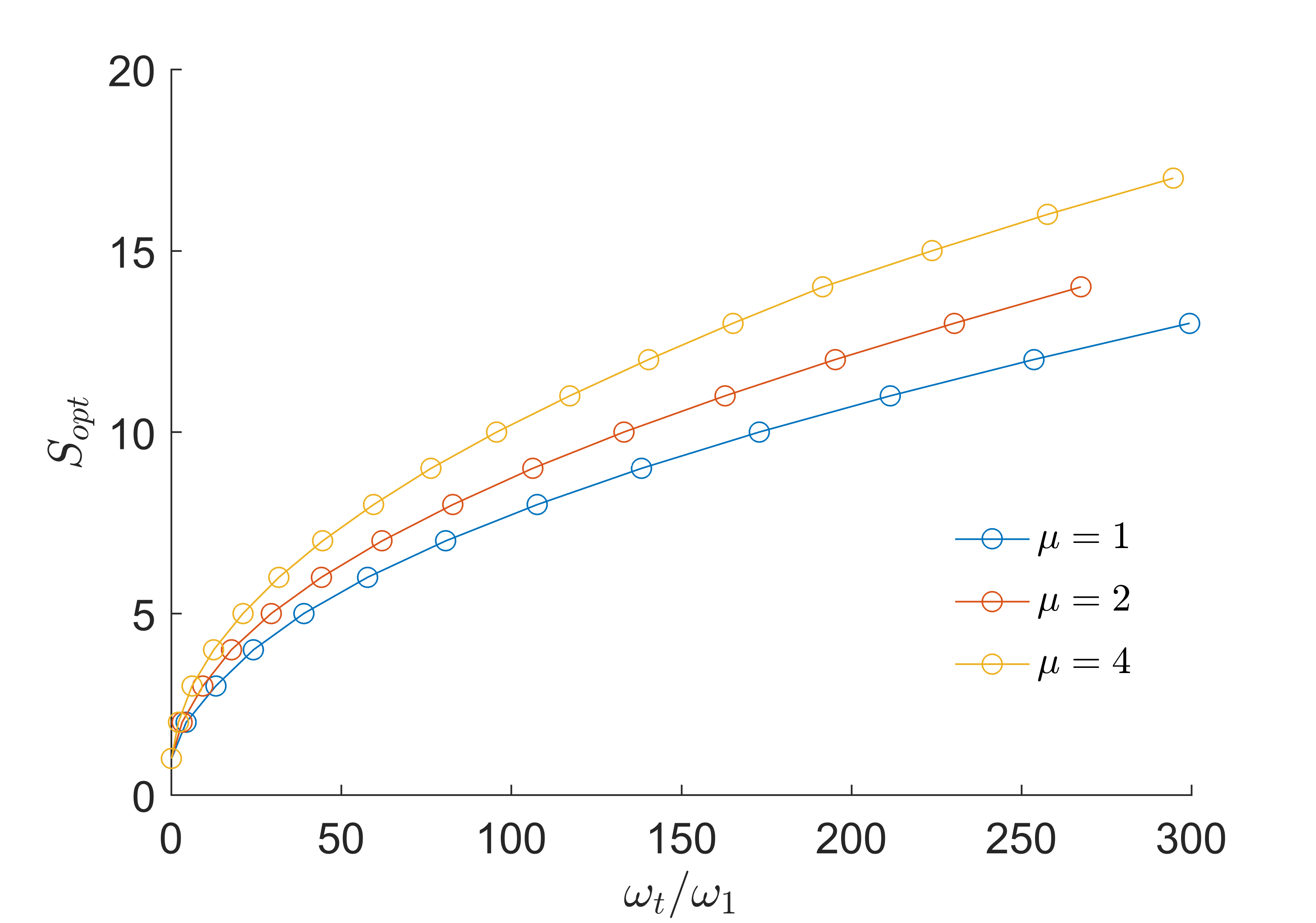}
	\caption{Number of resonators which maximizes effective bandgap width vs.\ target frequency for a uniform cantilever beam for three different values of $\mu$. The value of $S_{opt}$ for a given $\omega_t$ is given by taking the floor of the solid curve. The maximum values of $S_{opt}$ on each curve are valid until $\omega_t = 300\omega_1$. \label{fig:sopt_vs_wt}}
\end{figure}

The value $S_{opt}$ depends on the specific structure under consideration and the target frequency. This can be seen by maintaining a fixed mass ratio and finding $S_{opt}$ as a function of target frequency $\omega_t$. The resulting plot of $S_{opt}$ vs.\ $\omega_t$ is shown in Fig.\ \ref{fig:sopt_vs_wt} for three values of $\mu$. The number of attachments required to obtain the maximum bandgap width increases with increased target frequency for a fixed mass ratio. There is some dependence on $\mu$ that is only significant for very large changes in mass ratio. For a fixed target frequency, the number of attachment required to achieve the maximum bandgap width decreases with increased mass ratio. 

\begin{figure}
	\centering
	\includegraphics[]{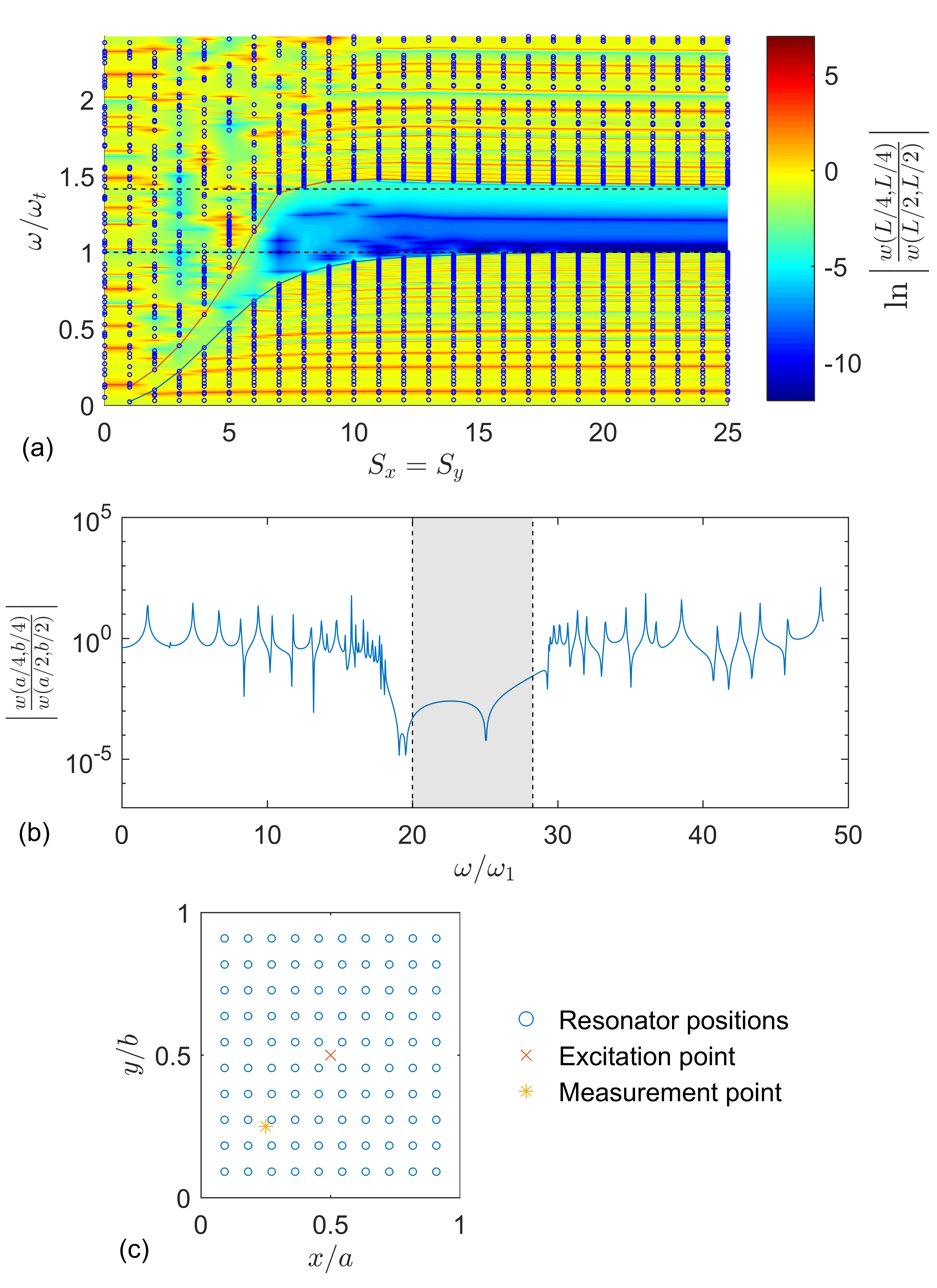}
	\caption{(a) Resonant frequencies and plate response at $(0.25a, 0.25b)$ for $\omega_t = 20\omega_1$, $\mu = 1$, $b = 1.1a$. Dashed lines show the bandgap predicted by Eq.\ \eqref{eq:bg}, solid lines track $\omega_{S+1}$ and $\omega_S$. (b) Plate response with $S_x = S_y = 10$. The dashed lines and gray region show the bandgap predicted by Eq.\ \eqref{eq:bg}. (c) Resonator positions, excitation location, and response measurement location for the results shown in (b). \label{fig:plate_figs}}
\end{figure}

These types of simulations can be extended to two-dimensional vibrating systems simply by changing the mode shapes and resonant frequencies for the system. For illustration, consider a thin rectangular isotropic plate simply supported on all edges with dimensions $a$ and $b$ in the $x$ and $y$ dimensions respectively, with $S_x$ resonators evenly spaced along the $x$ dimension and $S_y$ resonators evenly spaced along the $y$ dimension, such that the total number of resonators $S = S_xS_y$.  To reduce the number of independent variables, assume the resonators are uniform with mass ratio $\mu$ and target frequency $\omega_t$, and let $S_x = S_y$. With these assumptions, it is possible to solve for the resonant frequencies of the system as a function of $S_x$ only, along with the plate's response as a function of excitation frequency $\omega$. For simplicity, consider the response of the plate at $(x,y) = (0.25a, 0.25b)$ due to a point force at $(x,y) = (0.5a, 0.5b)$. The resulting resonances and plate response are shown in Fig.\ \ref{fig:plate_figs}a for $\mu = 1$, $\omega_t = 20\omega_1$. The plate's response is shown for $S_x = S_y = 10$ and the same parameters in Fig.\ \ref{fig:plate_figs}b, and the resonator positions are shown in Fig.\ \ref{fig:plate_figs}c. Again it is clear that, as the number of resonators becomes large, the bandgap converges to the frequency range given by Eq.\ \eqref{eq:bg}. Since the actual number of resonators is $S = S_x^2$, it is clear that significantly more resonators are required to create the bandgap for this system.

\subsection{Non-uniform Spatial Distribution of Resonators}
Since the locally resonant bandgap does not rely on any scattering phenomenon, there is no requirement that the resonators be evenly spaced on the structure. In the context of the infinite-absorber approximation Eq.\ \eqref{eq:infapprox}, it is sufficient only to have a large number of well-distributed resonators on the structure. Note that the accuracy of Eq.\ \eqref{eq:infapprox} depends on the mode shapes under consideration, so it is difficult to give a concrete numerical criterion on the required number of resonators for convergence. Furthermore, the modes that are relevant to the structure's response near the locally resonant bandgap depend on the target frequency of the resonators. 

\begin{figure}
	\centering
	\includegraphics[]{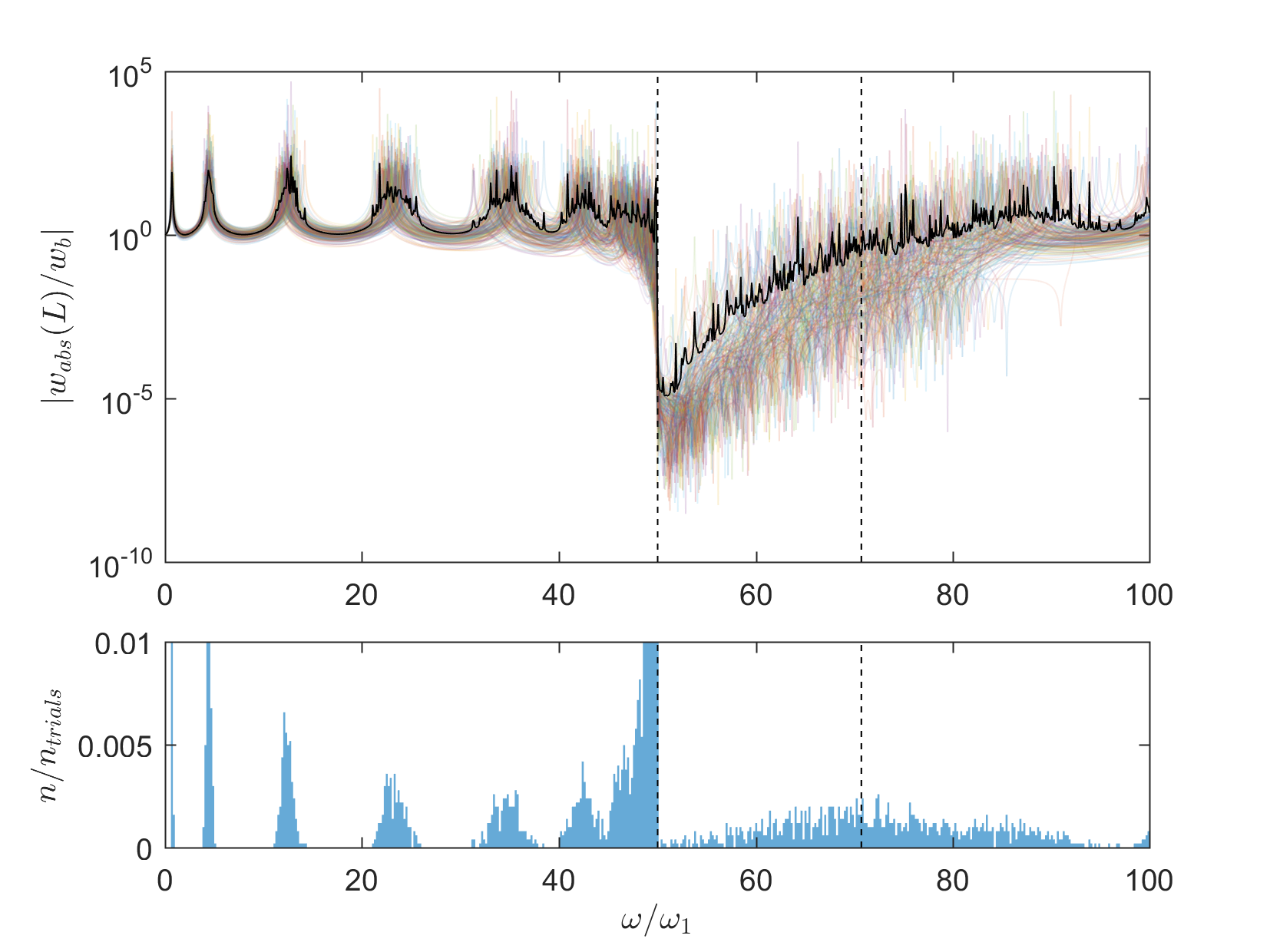}
	\caption{Top: Transmissibility plots for $n_{trials} = 200$ random arrangements of resonators, $S = 20$, $\omega_t = 50\omega_1$, $\mu = 1$. Solid black curve shows the average of all of the curves, vertical dashed lines show the ideal bandgap edge frequencies. Bottom: Histogram of the number of resonant frequencies $n$ from all of the random arrangements. The $y$-axis is truncated to emphasize the bandgap frequency range. \label{fig:rand_locations}}
\end{figure}

To quantify how non-uniform spatial distributions of resonators affect the locally resonant bandgap, consider a uniform cantilever beam with some non-uniform distribution of $S$ identical  resonators, each of them having identical mass and stiffness, hence identical natural frequency $\omega_t$, and overall mass ratio $\mu$. For each spatial arrangement, the resonant frequencies of the structure and the tip transmissibility can be calculated. This procedure can be repeated for $n_{trials}$ randomly generated arrangements, yielding the resonant frequencies and transmissibilities from every trial. One set of results from this procedure is shown in Fig.\ \ref{fig:rand_locations}. 

It is clear that the left edge of the bandgap is relatively insensitive to the specific arrangement of resonators, characterized by a sharp reduction in transmissibility in all cases. The right edge of the bandgap becomes significantly less well-defined, characterized by a wide frequency band where new resonances can be expected. Nevertheless, the average transmissibility curve (the solid black line) suggests that the vibration attenuation in the bandgap exists despite the occasional resonance appearing in that frequency range. Experimental results for non-uniform arrangements will be presented in Sec.\ \ref{subsec:nonuniform_exp}.

\subsection{Variation in Resonator Natural Frequencies}\label{subsec:param_unc}
\begin{figure*}
	\centering
	\includegraphics[]{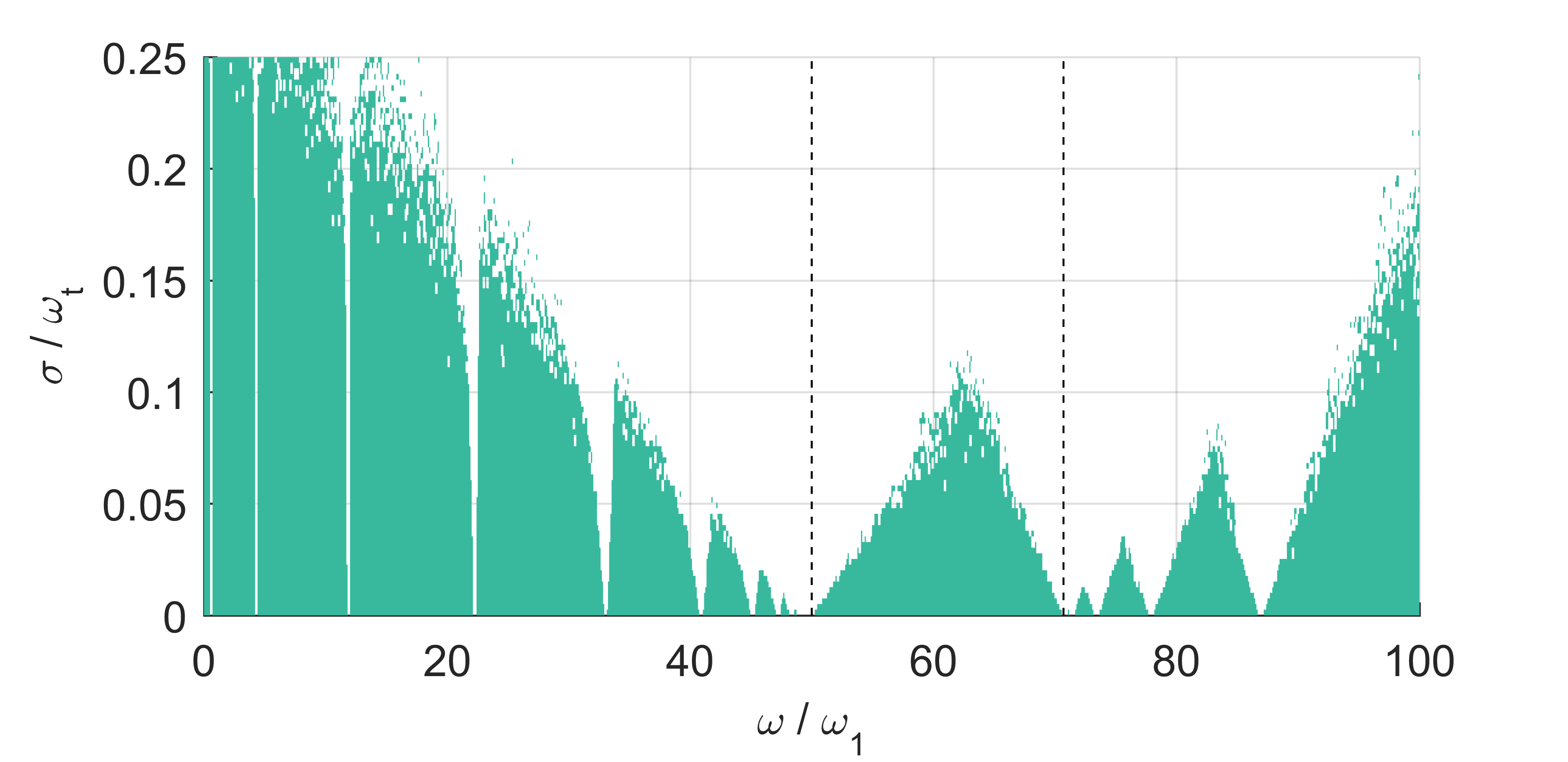}
	\caption{Plot showing frequencies where there were no resonances after $n_{trials} = 500$ at various parameter variations $\sigma$ with $S = 20$ uniformly distributed resonators, $\omega_t = 50\omega_1$, $\mu = 1$. The vertical dashed lines show the expected bandgap edge frequencies. \label{fig:wj_unc_zero}}
\end{figure*}

The derivation of the simple expression for the locally resonant bandgap edge frequencies, Eq.\ \eqref{eq:bg}, relies on the assumption that all of the resonators have the exact same target frequency $\omega_t$. This is a purely theoretical assumption; real resonators will never have exactly the same natural frequencies regardless of how well they are tuned. Manufacturing imperfections, the interface used in connecting the resonators to the main structure, or other factors will result in some small variation in the natural frequencies of the resonators. To quantify the effect of some small disorder in the resonator natural frequencies, consider a uniform spatial arrangement of resonators on the structure, with a normal random distribution of resonator natural frequencies, with mean $\omega_t$ and standard deviation $\sigma$. For a particular number of resonators $S$ and mass ratio $\mu$, consider $n_{trials}$ random sets of resonator natural frequencies. For each trial, the approximate resonant frequencies are obtained by solving the discretized eigenvalue problem, and the full set of resonant frequencies for all of the trials are obtained and shown as a histogram. This procedure is repeated at various standard deviations $\sigma$, yielding histograms for every value of $\sigma$. The resulting surface plot would show all of the resonant frequencies obtained from the $n$ trials vs.\ $\sigma$. In practice, the most useful piece of information from these studies is the points where there are $\emph{no}$ resonances, as would be expected in the ideal infinite-absorber bandgap. The surface plot showing only the points with no resonances is shown in Fig.\ \ref{fig:wj_unc_zero} for a uniform cantilever beam. 

With this type of graph, it is immediately apparent where the worst-case bandgap can be expected as the variation $\sigma$ increases. At $\sigma = 0$, there are no resonances throughout the entire bandgap, as expected. The bandgap becomes smaller as $\sigma$ increases, eventually disappearing completely. Note that the maximum allowable value of the normalized frequency variation $\sigma/\omega_t$ depends on the number of resonators and target frequency.

\section{Experimental Validation}

\begin{figure}
	\centering
	\includegraphics[width = 5in]{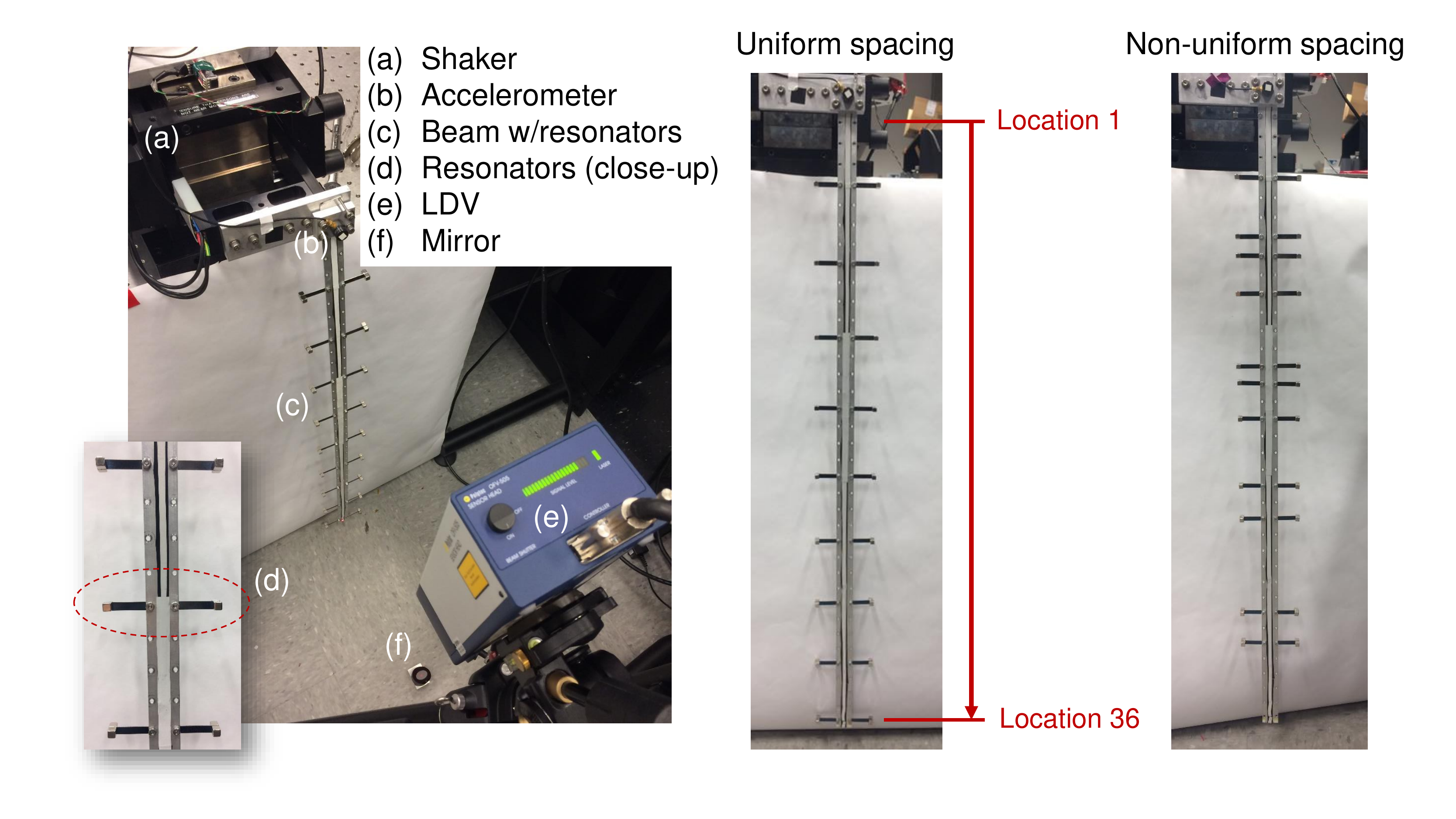}
	\caption{Left: Full experimental setup with $S = 9$ evenly spaced resonators. The LDV was oriented vertically at a $45^{\circ}$ mirror near ground level to direct the laser at the tip of the beam. Center: $S = 9$ evenly spaced resonators and the numbering scheme used to label the resonator positions. Location 1 is closest to the clamp and location 36 is closest to the free end of the beam. Right: Non-uniform spatial distribution of resonators (locations 4, 7, 8, 10, 14, 15, 17, 21, 23, 29, 31). \label{fig:expsetup}}
\end{figure}

To validate the model developed here, a locally resonant cantilever beam was built and tested. The experimental setup us shown in Fig.\ \ref{fig:expsetup}. The main beam consists of \SI{0.8}{mm} thick by \SI{2.54}{cm} wide by \SI{0.9}{m} long aluminum, with pairs of holes placed every \SI{2.54}{cm} along the length to allow for resonator placement. For resonators, spring steel cantilevers with tip masses were used. Each resonator consists of a piece of spring steel, \SI{0.5}{mm} thick by \SI{6.35}{mm} wide by \SI{8.26}{cm} long, clamped symmetrically on the beam via two screws and nuts, such that there is \SI{3.18}{cm} of spring steel extending past the main beam on both sides, forming two cantilevers. Two \SI{6.35}{mm} cube neodymium permanent magnets are placed at the tip of each cantilever, giving a total tip mass of \SI{3.6}{g}. The main beam was clamped vertically to an APS long stroke shaker, which excited the beam by base motion. The base acceleration was measured using an accelerometer, and the tip velocity was measured using a Polytec laser Doppler vibrometer (LDV), directed at the tip (free end of the beam) using a small mirror aligned at $45^\circ$ degrees relative to the floor.

The experiment was designed to allow flexibility in the placement of the resonators. Since it is expected that periodicity is not required under the approximation in Eq.\ \eqref{eq:infapprox}, tests were conducted using both uniform and non-uniform arrangements of resonators (Fig.\ \ref{fig:expsetup}). It is understood that in all of these tests, some small variation was present in the resonant frequencies of the attachments. An average value of $f_t = \SI{97.87}{Hz}$ was used for modeling purposes.

It is worth noting that the hardware required to attach the resonators to the beam adds a significant amount of mass to the system. Assuming the attachment hardware is small enough to act as a point mass, and assuming the point masses have a mass distribution proportional to the resonator mass distribution, placed at the same locations as the resonators $\vect{P}_j$, it can be shown that the adjusted mass ratio is
\begin{equation}
	\mu = \dfrac{\displaystyle\sum_{j=1}^S m_j}{\displaystyle\int_Dm(\vect{P})\dif D + \sum_{j=1}^Sm_{p,j}}
\end{equation}
where $m_{p,j}$ is the point mass (the mass of the clamping hardware) at $x = x_j$. For a finite number of attachments, point masses contribute to the mass matrix of the discretized system in a similar fashion to the resonators, without the frequency dependence. 

\subsection{Uniform Distribution of Resonators and Effect of Mass Ratio}\label{subsec:uniform_exp}
\begin{figure}
	\centering
	\includegraphics[]{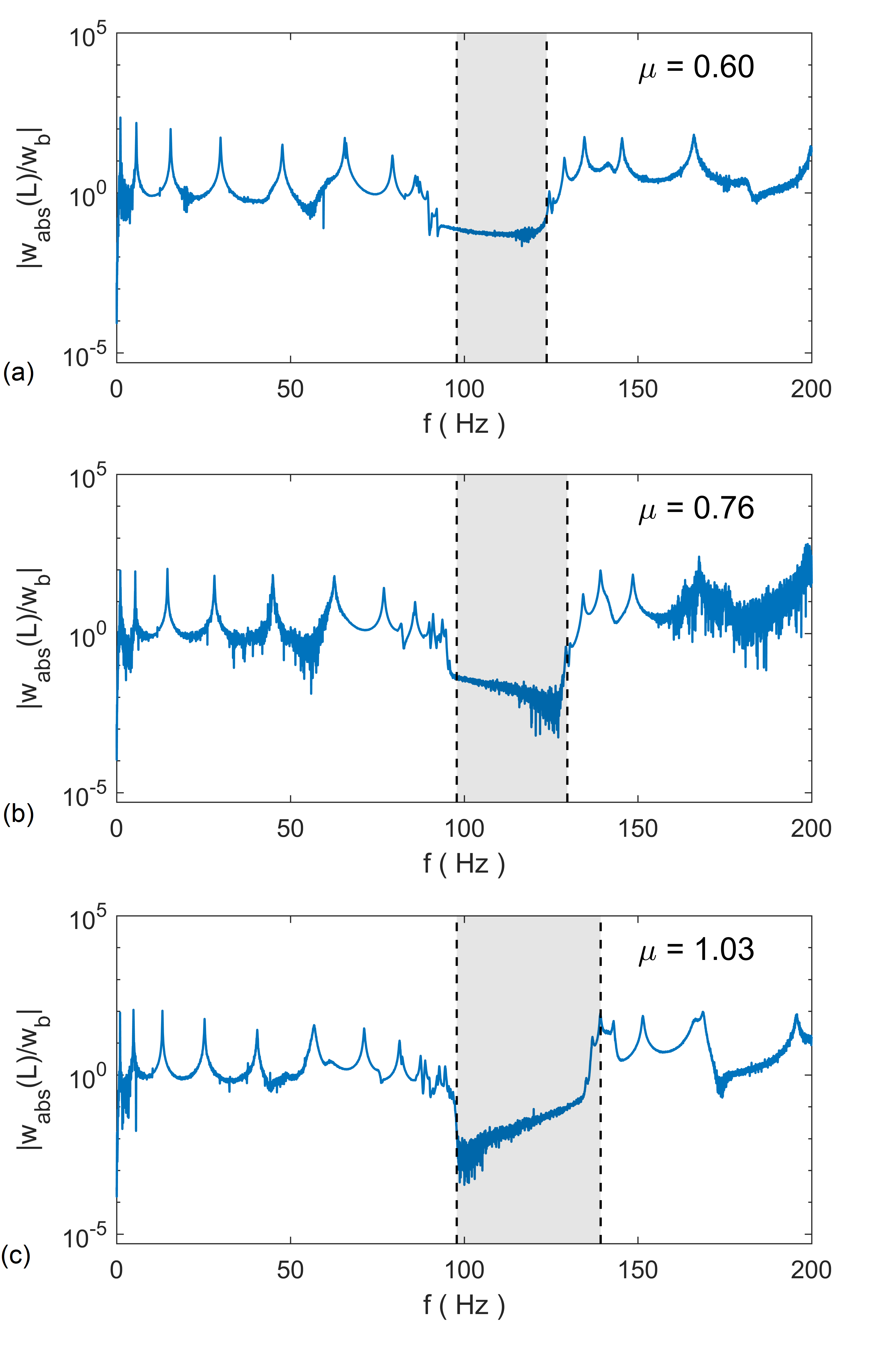}
	\caption{Experimental transmissibility plots for (a) $S = 9$, (b) $S = 12$, and (c) $S = 18$ identical uniformly distributed resonators showing the effect of increasing mass ratio from $\mu = 0.60$ to $\mu = 1.03$ from an increased number of resonators. Shaded regions give the expected bandgap frequency ranges for each mass ratio. \label{fig:uniform_spacing}}
\end{figure}

First we discuss the effect of mass ratio for the case of uniformly distributed resonators. The experimental setup was designed to permit multiple uniform arrangements of resonators, with 36 evenly spaced sets of holes where resonators could be attached. Tests were conducted using $S = 9$, $S = 12$, and $S = 18$ evenly spaced resonators. Although the tuning of the resonators was kept the same for each test, the mass ratio $\mu$ increased as the number of resonators increased. Thus, this set of tests provides validation for the expression for ideal bandgap edge frequencies in Eq.\ \eqref{eq:bg}. The results for the uniform spacing tests are shown in Fig.\ \ref{fig:uniform_spacing}.

From the experiment results, it is clear that the locally resonant bandgap is present for $S = 9$, $S = 12$, and $S = 18$ (yielding mass ratios of 0.60, 0.76, and 1.03 respectively), and that the bandgap edge frequencies predicted by Eq.\ \eqref{eq:bg} accurately represent the measured bandgaps. It is worth noting that only the mass of the beam, mass of the clamping hardware, tip masses, and attachment natural frequencies need to be measured to obtain the bandgap edge frequencies. A complete model for the structure is not necessary, as the bandgap edge frequencies are insensitive to the structure specific dynamics.

\subsection{Non-uniform Arrangements}\label{subsec:nonuniform_exp}
\begin{figure}
	\centering
	\includegraphics[]{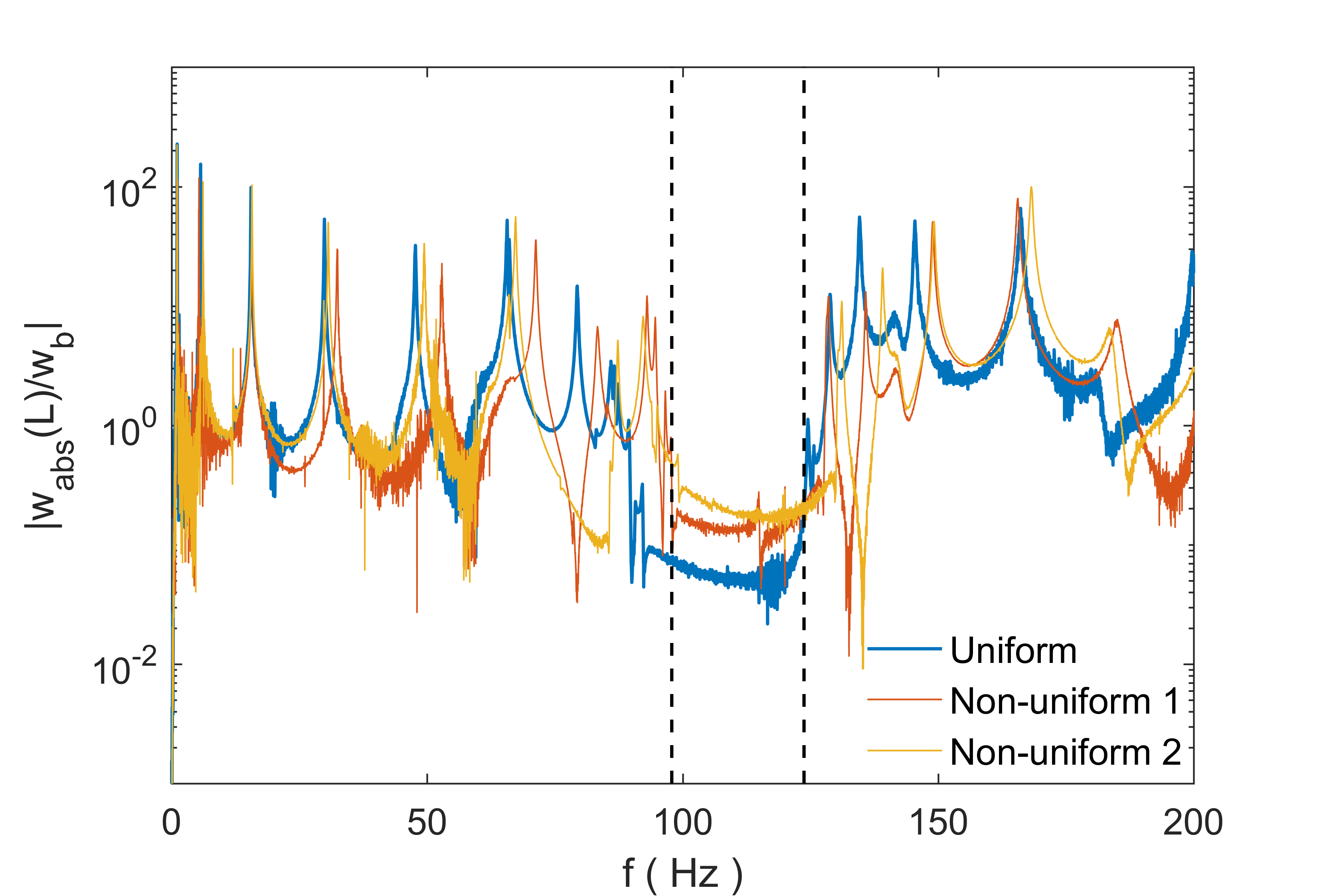}
	\caption{Experimental transmissibility for $S = 9$ uniformly distributed identical resonators and $S = 9$ non-uniformly placed identical resonators. The first non-uniform arrangement had resonators at locations 2, 9, 14, 15, 18, 19, 23, 25, and 36. The second non-uniform arrangement had resonators at locations 3, 6, 10, 17, 25, 26, 29, 32, and 36. The dashed lines show the expected bandgap edge frequencies.\label{fig:rand9}}
\end{figure}

Because the derivations developed in Sec.\ \ref{sec:amm} rely on the Riemann sum approximation in Eq.\ \ref{eq:infapprox}, rather than requiring a periodic arrangement of resonators, experiments were performed with non-uniform arrangements of resonators to determine if a bandgap could be created. Since the experimental setup permits only 36 discrete attachment points, each non-uniform arrangement was a random permutation of the integers 1 -- 36 with $S$ unique elements. Figure \ref{fig:rand9} shows a comparison between the uniform spacing FRF and two FRFs from non-uniform arrangements with $S = 9$ that appear to show the bandgap. In practice, it is likely simplest to use a uniform distribution of resonators to guarantee that the bandgap will appear. However, these results show that there is some additional flexibility in how locally resonant metamaterials can be designed without a periodic unit cell. 

\section{Conclusions}

In this paper, we have presented a general theory for bandgap estimation in general 1D or 2D vibrating finite structures using a differential operator formulation. Using the assumption of an infinite number of resonators on the structure tuned to the same frequency, a simple expression for the ``infinite-absorber" bandgap edge frequencies was derived. This expression applies to any canonical vibrating structure, whether 1D or 2D, has no dependency on the boundary conditions of the system, and depends only on the added mass ratio and target frequency. Unlike individual tuned mass absorbers, which must be tuned to a specific resonance of the system, the locally resonant bandgap can be placed in any frequency range regardless of the natural dynamics of the system. The locally resonant bandgap can be created in non-uniform structures so long as the resonator masses are distributed proportionally to the primary structure's mass distribution. This result could not be captured by widely-used unit-cell based dispersion analysis, which requires any spatially varying parameters to be periodic. 

For locally resonant metamaterials with finite dimensions and a finite number of resonators, the key approximation required for the bandgap to appear is a Riemann sum approximation of an integral. Thus, some minimum number of resonators is required for the bandgap to appear, which strongly depends on what modal neighborhood is being targeted. More resonators are necessary when targeting a high modal neighborhood; this type of mode shape dependent information could not be captured with dispersion analysis of an individual unit cell.  Furthermore, it was observed that there is an optimal number of resonators for the widest possible bandgap, which is larger than the infinite-absorber bandgap. The optimal number of resonators depends on the specific structure under consideration, and strongly depends on the modal neighborhood being targeted. At a fixed mass ratio, the optimal number of resonators increases with target frequency. Similarly, at a fixed target frequency, the optimal number of resonators increases with mass ratio. As the number of resonators becomes large, the bandgap clearly converges to the infinite-absorber bandgap. The infinite-absorber bandgap edge frequencies were also validated with the well-known plane wave expansion method. 

Numerical studies show that the bandgap width is reduced by variation in the resonant frequencies of the resonators, but can tolerate some small variation that depends on the number of resonators and modal neighborhood of the bandgap. Additionally, there is no requirement for periodicity in resonator placement, as non-uniform spatial distributions of resonators are capable of creating the bandgap. The bandgap expression and the effect of changing mass ratio were validated experimentally with a cantilever beam with small spring steel cantilevers with tip masses as resonators. The robustness of the bandgap in case of non-periodic spatial arrangement of the resonators was also shown experimentally.  

\section{Acknowledgments}
This work was supported by the Air Force Office of Scientific Research grant FA9550-15-1-0397 ``Integrated multi-field resonant metamaterials for extreme, low frequency damping" monitored by Dr.\ B.L.\ Lee.

\appendix
\section{Governing equations for systems excited by base motion}\label{sec:base_motion}
For a structure excited by some base displacement $w_b(t)$ with relative vibration $w(\vect{P},t)$, the governing equations in physical coordinates are
\begin{equation}
\mathcal{L}w(\vect{P},t) + m(\vect{P})\ddot{w}(\vect{P},t) - \sum_{j=1}^Sm_j\omega_{a,j}^2u_j(t)\delta(\vect{P} - \vect{P}_j)  = f(\vect{P},t) - m(\vect{P})\ddot{w}_b(t), \quad \vect{P} \in D \label{eq:ge_wb}
\end{equation}
\begin{equation}
m_j\ddot{u}_j(t) + m_j\omega_{a,j}^2u_j(t) + m_j\ddot{w}(\vect{P}_j, t) = -m_j\ddot{w}_b(t)
\end{equation}
In modal coordinates, these become
\begin{equation}
	\ddot{\eta}_r(t) + \omega_r^2\eta_r(t) + \sum_{j=1}^Sm_j\phi_r(\vect{P}_j)\sum_{k=1}^N\ddot{\eta}_k(t)w_k(\vect{P}_j) + \sum_{j=1}^Sm_j\ddot{u}_j(t)\phi_r(\vect{P}_j) = q_r(t) 
\end{equation}
\begin{equation}
m_j\ddot{u}_j(t) + m_j\omega_{a,j}^2u_j(t) + m_j\sum_{r=1}^N\ddot{\eta}_r(t)\phi_r(\vect{P}_j) = -m_j\ddot{w}_b(t)
\end{equation}
where
\begin{equation}
q_r(t) = \int_Df(\vect{P},t)\phi_r(\vect{P})\dif D - \ddot{w}_b(t)\left(\int_Dm(\vect{P})\phi_r(\vect{P})\dif D + \sum_{j=1}^Sm_j\phi_r(\vect{P}_j)\right)
\end{equation}
The remaining procedure is the same, yielding a system of equations in the Laplace domain of the same form as Eq.\ \eqref{scoupled}, where 
\begin{equation}
Q_r(s) = \int_DF(\vect{P},s)\phi_r(\vect{P})\dif D - W_bs^2\left(\int_Dm(\vect{P})\phi_r(\vect{P})\dif D + \frac{\omega_t^2}{s^2 + \omega_t^2}\sum_{j=1}^Sm_j\phi_r(\vect{P}_j)\right)
\end{equation}
The remaining procedure is the same as shown in Sec.\ \ref{sec:amm}.

\section{Stiffness operators and mass densities for canonical vibrating systems}\label{sec:lmdefs}
\subsection{Strings, Rods, and Shafts}
The transverse vibration of strings, longitudinal vibration of rods, and torsional vibration of shafts all have second order stiffness differential operators ($p = 1$), and so they can all be treated in a similar fashion. The stiffness operator has the form
\begin{equation}
\mathcal{L} = -\dod[]{}{x}\left[K(x)\dod[]{}{x}\right]
\end{equation}
where
\begin{equation}
K(x) = \begin{cases}
T(x) & \text{strings (tension)}\\
EA(x) & \text{rods (longitudinal stiffness)}\\
GJ(x) & \text{shafts (torsional stiffness)}
\end{cases}
\end{equation}
For strings and rods, $m(x)$ is the mass per unit length. For shafts, $m(x)$ is the rotary inertia about the center of mass per unit length. 
\subsection{Beams}\label{subsec:beams}
For a slender beam governed by the Euler-Bernoulli beam theory, the stiffness operator is fourth order ($p = 2$), given by
\begin{equation}
\mathcal{L} = \dod[2]{}{x}\left[EI(x)\dod[2]{}{x}\right]
\end{equation}
where $EI(x)$ is the bending stiffness of the beam. The mass density $m(x)$ for beams is the mass per unit length of the beam. 

Since the numerical studies presented Sec.\ \ref{sec:numstudies} focus on a cantilever beam, the appropriate mode shapes and eigenvalues are presented here. For a uniform cantilever beam of length $L$, bending stiffness $EI$, and mass per unit length $m$, the mass-normalized mode shapes of the beam are
\begin{equation}
	\phi_r(x) = \frac{1}{\sqrt{mL}}\left[\cos\left(\dfrac{\lambda_r x}{L}\right) - \cosh \left(\dfrac{\lambda_r x}{L}\right) + \left(\dfrac{\sin\lambda_r - \sinh\lambda_r}{\cos\lambda_r + \cosh\lambda_r}\right)\left(\sin\left(\dfrac{\lambda_r x}{L}\right) - \sinh\left(\dfrac{\lambda_r x}{L}\right)\right)\right]\label{mode_cantilever}
\end{equation}
where $\lambda_r$ is the $r$th positive, real solution to the characteristic equation
\begin{equation}
\cos\lambda \cosh \lambda + 1 = 0\label{res_cantilever}
\end{equation} 
Note that it is necessary only to solve for the first several solutions to this equation numerically, as larger solutions are well-approximated by
\begin{equation}
\lambda_r \approx \dfrac{(2r - 1)\pi}{2}, \quad r > 5 
\end{equation}
The resonant frequencies are given by
\begin{equation}
\omega_r = \lambda_r^2 \sqrt{\dfrac{EI}{mL^4}}
\end{equation}
Note that due to the hyperbolic functions in Eq.\ \eqref{mode_cantilever}, it is necessary to use an approximate mode shape for $r > 10$ to avoid numerical issues, given by
\begin{equation}
\phi_r(x) \approx \frac{1}{\sqrt{mL}}\left[\cos\left(\dfrac{\lambda_r x}{L}\right) - \sin\left(\dfrac{\lambda_r x}{L}\right) - e^{-\lambda_rx/L} - e^{\lambda_r(x/L - 1)}\sin\lambda_r\right]
\end{equation}
For a uniform cantilever beam experiencing base motion $w_b(t)$ with relative transverse vibration $w(x,t)$, the modal excitation can be written as
\begin{equation}
Q_r(s) = -s^2\left(m\int_0^L\phi_r(x)\dif x + \frac{\omega_t^2}{s^2 + \omega_t^2}\sum_{j=1}^Sm_j\phi_r(x_j)\right)
\end{equation}
Note that
\begin{equation}
\lim_{S\rightarrow\infty}Q_r(s) = -ms^2\left(1 + \frac{\mu\omega_t^2}{s^2 + \omega_t^2}\right)\int_0^L\phi_r(x)\dif x
\end{equation}

\subsection{Membranes}
For a uniform membrane with constant tension, the stiffness operator is
\begin{equation}
\mathcal{L} = -T\nabla^2
\end{equation}
where $\nabla^2$ is the Laplacian and $T$ is the tension per length in the membrane. The mass density $m(\vect{P})$ is the mass per unit area of the membrane. 

\subsection{Plates}\label{subsec:plates}
For a uniform thin plate governed by Kirchhoff-Love plate theory, the stiffness operator is
\begin{equation}
\mathcal{L} = D_E\nabla^4
\end{equation}
where $\nabla^4$ is the biharmonic operator,
\begin{equation}
D_E = \dfrac{Eh^3}{12(1 - \nu^2)}
\end{equation}
is the flexural rigidity, $E$ is the Young's modulus of the material, $\nu$ is the Poisson's ratio of the material, and $h$ is the thickness of the isotropic plate. The mass density $m(\vect{P})$ is the mass per unit area of the plate. For a uniform rectangular plate of dimensions $a$ and $b$ in the $x$ and $y$ axes respectively, simply supported (pinned) on all edges, the mass-normalized mode shapes are
\begin{equation}
\phi_{rs}(x,y) = \frac{2}{\sqrt{m a b}}\sin\left(\frac{r\pi x}{a}\right)\sin\left(\frac{s\pi y}{b}\right)
\end{equation}
with corresponding resonant frequencies
\begin{equation}
\omega_{rs} = \pi^2\left[\left(\frac{r}{a}\right)^2 + \left(\frac{s}{b}\right)^2\right]\sqrt{\frac{D_E}{m}}
\end{equation}
For a point force excitation $F_o(t)$ at a point $(x_f, y_f)$, the modal excitation is
\begin{equation}
q_{rs}(t) = F_o(t)\phi_{rs}(x_f,y_f)
\end{equation}
\section{Plane Wave Expansion Method Using Operator Formulation}\label{sec:pwem_app}

Consider the governing equation of the general vibrating system given in operator notation under harmonic excitation
\begin{gather}
\mathcal{L}\left[w_1(\vect{P})\right] - \omega^2m(\vect{P})w_1(\vect{P}) = -\sum_{j}k_j(w_1(\vect{P}_j) - w_2(\vect{P}_j))\delta(\vect{P} - \vect{P}_j), \quad \vect{P} \in D\\
-\omega^2m_jw_2(\vect{P}_j) = k_j(w_1(\vect{P}_j) - w_2(\vect{P}_j))
\end{gather}
where $\mathcal{L}$ is the linear, self-adjoint stiffness operator of order $2p$, $w_1(\vect{P})$ is the displacement amplitude of the point $\vect{P}$, $\omega$ is the excitation frequency, $m(\vect{P})$ is the mass density of the structure, $k_j$ is the stiffness of the $j$th resonator, $\vect{P}_j$ is the attachment point of the $j$th resonator, $w_2(\vect{P})$ is the absolute displacement of the resonator at point $\vect{P}$, and $\delta(\vect{P})$ is the spatial Dirac delta function. For a uniform structure, the stiffness operator $\mathcal{L}$ has the form
\begin{equation}
\mathcal{L} = \mathcal{L}\left(\dpd{}{x}, \dpd{}{y}\right) = P_\mathcal{L}\left(\dpd{}{x}, \dpd{}{y}\right)
\end{equation}
where $P_\mathcal{L}$ is a polynomial of order $2p$. For 1D systems, the polynomial is in only a single variable. Following the general plane wave expansion procedure, it can be shown that the governing equations for the system become
\begin{gather}
\Delta D_j W_1(\vect{G}_n)P_\mathcal{L}\left(-i(\vect{k} + \vect{G}_n)\right) - \Delta D_j \omega^2mW_1(\vect{G}_n) = -k_j\left(\sum_m W_1(\vect{G}_m) - w_2(\vect{0})\right)\\
-\omega^2m_jw_2(\vect{0}) = k_j\left(\sum_mW_1(\vect{G}_m) - w_2(\vect{0})\right)
\end{gather}
where $\vect{G}_m$ are the reciprocal lattice vectors, $k$ is the Bloch wavevector,  $W_1(\vect{G}_m)$ is the plane wave amplitude corresponding to reciprocal lattice vector $\vect{G}_m$, and $\Delta D_j$ is the length or area of the unit cell. The resulting eigenvalue problem must be solved at every Bloch vector $k$ in the irreducible region of the first Brillouin zone. If the index $m$ is taken to go from $-M$ to $M$, with $N = 2M+1$ total plane waves used in the expansion, the eigenvalue problem is order $N+1$ (one-dimensional systems) or $N^2+1$ (two-dimensional systems). The appropriate parameters and polynomials $P_\mathcal{L}$ for the canonical vibrating systems discussed in \ref{sec:lmdefs} are given in Table \ref{table:pwem}.

\begin{table}
	\centering
	\caption{Relevant unit cell dimensions, polynomials, and mass densities for the canonical vibrating systems for the plane wave expansion method. The variables $x, y$ refer to the $x$ and $y$ components of the input to $P_\mathcal{L}$.\label{table:pwem}}
	\begin{tabular}{c|c|c|c}
		{} &	$\Delta D_j$ & $P_\mathcal{L}$ & $m(\vect{P})$\\\hline\hline
		Strings & length of unit cell & $Tx^2$ & mass per unit length\\
		Rods & length of unit cell & $EAx^2$ & mass per unit length\\
		Shafts & length of unit cell & $GJx^2$ & rotary inertia per unit length\\
		Beams & length of unit cell & $EIx^4$ & mass per unit length\\
		Membranes & area of unit cell & $T(x^2 + y^2)$ & mass per unit area\\
		Plates & area of unit cell & $D_E(x^2+y^2)^2$ & mass per unit area
	\end{tabular}
\end{table}

\section*{References}

\bibliography{modal_bandgap_general}

\end{document}